\documentclass[aps,pra,twocolumn,showpacs,superscriptaddress]{revtex4}
\usepackage{graphicx}
\usepackage{amsmath}
\usepackage{amssymb}
\usepackage{epstopdf}
\usepackage{amsfonts}
\usepackage{dcolumn}
\usepackage{bigints}

\begin{document}
\title{Magic Conditions for Multiple Rotational States of Bialkali Molecules in Optical Lattices}
\author{Q. Guan}
\affiliation{Department of Physics, Temple University, Philadelphia, PA 19122, USA}
\author{Simon L. Cornish}
\affiliation{Joint Quantum Centre (JQC) Durham-Newcastle, Department of Physics, Durham University, South Road, Durham, DH1 3LE}
\author{S. Kotochigova}
\affiliation{Department of Physics, Temple University, Philadelphia, PA 19122, USA}

\date{\today}

\begin{abstract}
We investigate magic-wavelength trapping of ultracold bialkali molecules in the vicinity of weak optical transitions from the vibrational ground state of the X$^1\Sigma^+$ potential to low-lying rovibrational states of the b$^3\Pi_0$ potential, focussing our discussion on the $^{87}$Rb$^{133}$Cs molecule in a magnetic field of $B=181\,$G. We show that a frequency window exists between two nearest neighbor vibrational poles in the dynamic polarizability where the trapping potential is ``near magic'' for multiple rotational states simultaneously. We show that the addition of a modest DC electric field of $E=0.13\,\text{kV}/\text{cm}$ leads to an exact magic-wavelength trap for the lowest three rotational states at a angular-frequency detuning of $\Delta_{v'=0} = 2\pi\times 218.22$\,GHz from the X$^1\Sigma^+ (v=0, J=0)\rightarrow$ b$^3\Pi_0 (v'=0, J=1)$ transition. We derive a set of analytical criteria that must be fulfilled to ensure the existence of such magic frequency windows and present an analytic expression for the position of the frequency window in terms of a set of experimentally measurable parameters. These results should inform future experiments requiring long coherence times on multiple rotational transitions in ultracold polar molecules.

\end{abstract}
\maketitle

\section{Introduction}
\label{sec_introduction}
Ultracold polar molecules present a wealth of opportunities in quantum science and technology~\cite{Carr:2009}. Proposed applications span the fields of precision measurement and metrology~\cite{Zelevinsky:2008,Salumbides:2011,Salumbides:2013,Tarbutt:2013,Schiller:2014,Borkowski:2018,Borkowski:2019}, quantum-state resolved chemistry~\cite{Krems:2008,Bell:2009,Ospelkaus:2010,Dulieu:2011,Balakrishnan:2016}, dipolar quantum matter~\cite{Santos:2000,Micheli:2007,Pollet:2010,Capogrosso_Sansone:2010,Baranov:2012,Lechner:2013}, quantum simulation~\cite{Barnett:2006,Micheli:2006,Buchler:2007,Macia:2012,Manmana:2013,Gorshkov:2013} and quantum information processing~\cite{Demille:2002,Yelin:2006,Zhu:2013,Herrera:2014,Ni:2018,Sawant:2020,Hughes:2019}. Recent experimental progress on the production of ultracold molecules by association~\cite{Ni:2008,Danzl:2008,Lang:2008,Takekoshi:2014,Molony:2014,Park:2015,Guo:2016,Rvachov:2017,Seesselberg:2018b,Yang:2019,Voges:2020} and direct laser cooling~\cite{Shuman:2010,Barry:2014,Truppe:2017,Kozyryev:2017,Anderegg:2018,Collopy:2018} has brought many of these applications within reach.

In the realm of quantum simulation and computation, the rotational structure of ultracold molecules provides a rich basis of long-lived states in which to encode pseudo-spins or quantum information. Owing to the permanent molecular-frame electric dipole moment, the rotational states can be conveniently manipulated with microwave fields, as already demonstrated in a number of settings~\cite{Ospelkaus:2010b,Yan:2013,Gregory:2016,Will:2016,Guo:2018,Blackmore:2020}. Moreover, laboratory-frame dipole moments can be engineered using applied electric fields or superpositions of rotational states. The resulting long-range interaction between molecules can be exploited to realise model Hamiltonians in quantum magnetism~\cite{Barnett:2006,Micheli:2006,Gorshkov:2011,Gorshkov:2011b,Manmana:2013,Hazzard:2013} and two-qubit gates for quantum information processing~\cite{Demille:2002,Yelin:2006,Zhu:2013,Herrera:2014,Ni:2018,Sawant:2020,Hughes:2019}. To generate useful interaction strengths necessitates inter-molecular distances below a micrometre. This is most readily achieved using optical potentials, either in the form of an optical lattice~\cite{Moses:2015,Reichsollner:2017} or an array of optical tweezers~\cite{Liu:2019,Anderegg:2019}.

For diatomic molecules, such as ground-state bialkali molecules~\cite{Ni:2008,Danzl:2008,Lang:2008,Takekoshi:2014,Molony:2014,Park:2015,Guo:2016,Rvachov:2017,Seesselberg:2018b,Yang:2019,Voges:2020}, the dynamic polarizability along the molecular axis ($\alpha_{\parallel}$) is, in general, different from that perpendicular to it ($\alpha_{\perp}$). For light polarized at an angle $\theta$ to the molecular axis, this leads to a dynamic polarizability in the body-fixed frame given by,
\begin{align}
\label{eqn0}
\alpha(\theta) = \alpha^{(0)} + \alpha^{(2)} P_2(\cos(\theta)),
\end{align}
where $\alpha^{(0)}=\frac{1}{3}(\alpha_{\parallel}+2\alpha_{\perp})$ and $\alpha^{(2)}=\frac{2}{3}(\alpha_{\parallel}-\alpha_{\perp})$ are the isotropic and anisotropic components of the polarizability tensor, respectively. $\alpha_{\parallel}$ and $\alpha_{\perp}$ result from a sum over all allowed molecular transitions for the component of the dipole operator parallel or perpendicular to the molecular axis, respectively, and are smooth functions of wavelength in the regime where the frequency of the trapping laser is far-detuned from any rovibronic transitions~\cite{Kotochigova:2006,Vexiau:2017,Li:2017}. In the lab frame, the dynamic polarizability can be thought of as the spatial average of $\alpha(\theta)$. Although $\alpha^{(0)}$ is the same for all rotational states, $\alpha^{(2)}$ strongly mixes states with different rotational projections in excited rotational states. It follows that for molecules confined in an optical potential, the anisotropic polarizability leads to rotational transition frequencies that are strongly dependent on the intensity and polarization of the trapping light. The concomitant state-dependent light shifts make it highly challenging to achieve rotational coherence times that are sufficiently long to be sensitive to the $\sim$kHz interaction strengths~\cite{Yan:2013,Seesselberg:2018} typical of most molecules. Nevertheless, several approaches have been developed to match the polarizabilities of two specific states within a molecule. These include judicious choice of the intensity and polarisation of the trapping light~\cite{Neyenhuis:2012,Gregory:2017,Blackmore:2018} and the addition of applied electric fields to simplify the couplings within the molecule~\cite{Kotochigova:2010,Li:2017,Seesselberg:2018}.

Inspired by the magic-wavelength traps used in atomic clocks~\cite{Katori:2003,Ye:2008}, it is natural to investigate magic-wavelength trapping for molecules. Intuitively, magic trapping independent of the molecular rotational state can be realized under the condition of $\alpha^{(2)}=0$. To search for this condition, one needs to tune the trapping laser wavelength into a regime where there is significant interplay between several ro-vibrational poles in $\alpha_{\parallel}$ and $\alpha_{\perp}$. Indeed, following this approach, very recent work has demonstrated state-insensitive trapping for two vibrational~\cite{Kondov:2019} or rotational~\cite{Bause:2019} levels. These magic-frequency traps show reduced sensitivity to experimental parameters, enabling longer coherence times to be achieved. However, numerous proposed applications make greater use of the rich internal structure of molecules by simultaneously addressing \textit{more than two} rotational levels. Examples include, coupling three rotational levels with microwave fields to realize highly tunable models in quantum magnetism~\cite{Gorshkov:2011b} and mapping many rotational levels onto a synthetic dimension~\cite{Sundar:2018}. It is therefore pertinent to ask whether the concept of a magic frequency trap can be extended to multiple rotational levels simultaneously.

In this work, we investigate magic-wavelength trapping of ultracold bialkali molecules in the vicinity of weak optical transitions from the vibrational ground state of the X$^1\Sigma^+$ potential to low-lying rovibrational states of the b$^3\Pi_0$ potential, focussing our discussion on the $^{87}$Rb$^{133}$Cs molecule. We show that a magic trapping frequency window for multiple rotational states of the X$^1\Sigma^+$ potential exists between two nearest neighbor vibrational poles of the b$^3\Pi_0$ potential, far away from any rotational poles. Within this window, the laser trapping is ``near magic" for multiple rotational states simultaneously and is exactly magic for pairs of neighboring rotational states at specific laser frequencies. Moreover, the ``near magic" frequency window can be tuned to a true magic frequency for the lowest three rotational states by applying an experimentally accessible DC electric field. This true triple magic condition is expected to be useful for future studies of synthetic spin-1 systems using ultracold molecules. The existence of such a magic frequency window relies on a set of strict criteria which we derive analytically. We show that these criteria can be satisfied near the narrow X$^1\Sigma^+\rightarrow\mathrm{b}^3\Pi_0$ transitions for heavy molecules, including $^{87}$Rb$^{133}$Cs and $^{23}$Na$^{87}$Rb. We also derive an analytic expression for the position of the frequency window in terms of a set of experimentally measurable parameters, such as transition widths and transition wavelengths. This will provide a straightforward, self-consistent approach to search for the magic trapping frequency window in future experiments.

This paper is organized as follows. Section~\ref{hamiltonian} presents the general theoretical framework describing the molecular rotational states in the lowest vibrational state of the ground electronic potential in the presence of applied magnetic, electric and optical fields. In section~\ref{spectrum}, we discuss the hyperfine structure of the $^{87}$Rb$^{133}$Cs molecule in the presence of applied magnetic and electric fields with a view to identifying the best target states in each rotational level for magic trapping. In section~\ref{ac_stark}, we consider the AC-Stark shift and dynamic polarizability of $^{87}$Rb$^{133}$Cs molecules in the vicinity of the weakly allowed X$^1\Sigma^+\rightarrow\mathrm{b}^3\Pi_0$ transitions. In section~\ref{magic}, we identify magic trapping frequencies by searching for crossings among the frequency-dependent dynamic polarizability curves of different rotational states. We present a simple analytic treatment that shows excellent agreement with our numerical results, both near-resonance and in the magic frequency window between two vibrational poles. Imaginary polarizabilities for rotational states in the magic frequency window are also calculated. In section~\ref{discussion}, we discuss the wider significance of our work, before concluding in section~\ref{conclusion}.

\section{Theoretical Framework}
\label{hamiltonian}
We focus on the molecular rotational states $\vec J$ associated with the $v=0$ vibrational state of the ground electronic state of RbCs.
The effective Hamiltonian that describes the system in the presence of a static magnetic field  $\vec{B}$, a static electric field  $\vec{E}$, and an optical laser field of intensity $I$ \cite{APetrov:2013, Gregory:2016, Li:2017} is given by:
\begin{eqnarray}
\label{h_eff}
H = H_\text{rot} + H_Z +  H_\text{hf} + H_\text{DC} + H_\text{AC},
\end{eqnarray}
where the rotational Hamiltonian is
\begin{eqnarray}
\label{rot}
H_\text{rot} =  B_v \vec{J}^2\,,
\end{eqnarray}
the Zeeman Hamiltonian is
\begin{eqnarray}
\label{zeeman}
H_Z = - g_r \mu_{\rm N}  \vec{J}\cdot \vec{B} - \sum_{k=1}^2 g_k \mu_{\rm N} \vec{I}_k \cdot \vec{B}(1-\sigma_k)\,,
\end{eqnarray}
the nuclear quadrupole interaction is
\begin{eqnarray}
\label{hyperfine}
H_{\text{hf}} = \sum_{k=1}^2 \frac{(eqQ)_k}{I_k(I_k-1)}C_2(\alpha,\beta) T_2 (\vec{I}_k, \vec{I}_k)\,,
\end{eqnarray}
and the DC-Stark shift is
\begin{eqnarray}
\label{dc}
H_\text{DC} = -\vec{d}\cdot\vec{E}\,.
\end{eqnarray}
In Eqs~\eqref{rot}-\eqref{dc}  $\vec{J}$, $\vec{I}_k$, and $\vec{d}$ denote the molecule orbital angular momentum operator, the nuclear spin operators for the $k$-th atom, and the permanent molecular electric dipole moment operator, respectively.
The nuclear quadrupole interaction $H_{\text{hf}}$ couples the nuclear spin to rotational states and depends on the  quadrupole coupling constants $(eqQ)_k$ for Rb and Cs obtained from Refs~\cite{Gregory:2016}.
The operator $T_2 (\vec{I}_k, \vec{I}_k)$ is a rank-2 tensor and $C_2(\alpha,\beta) = \sqrt{4\pi/5}Y_{20}(\alpha,\beta)$ is the modified spherical harmonic  function, where the angles $\alpha$, $\beta$ describe the orientation of the diatomic molecule in the space-fixed coordinate frame. In these equations $B_v$ is the rotational constant, $\mu_{\rm N}$ is the  nuclear magneton, and
$g_r$ is the molecule rotational $g$-factor. Moreover, $g_k$ and $\sigma_k$ with $k=1,2$ are  nuclear-spin $g$-factors and  isotropic molecular nuclear shielding factors, respectively.

Here, the direction of the external magnetic field is our quantization axis along which we define  projection quantum numbers of angular momenta. The matrix elements of the Hamiltonian are determined in low-energy  set of basis functions  $|J, M; m_1, m_2\rangle$ ,
where $J$ and $M$ are the orbital angular momentum and  its associated projection, respectively.
Quantum numbers $m_k$ are nuclear spin projections of the $k$-th atom.

The AC-Stark Hamiltonian $H_\text{AC}$ in Eq.~\eqref{h_eff} is constructed up to second order in the electric field strength of the
driving laser in the regime where the AC-Stark shift is much smaller than the rotational constant. In this regime, the AC-Stark Hamiltonian $H_\text{AC}$ is
\begin{eqnarray}
H_{\text{AC}} &=&- \frac{I}{\epsilon_0 c}
\sum_{\substack{J, M, M', \\  m_1, m_2}}
|J, M'; m_1, m_2\rangle\langle J, M; m_1, m_2|\nonumber\\
&& \quad\times
\sum_f\frac{\langle J, M'|\vec{d}_{\rm tr}\cdot\vec{\epsilon}^*|f\rangle\langle f|\vec{d}_{\rm tr}\cdot\vec{\epsilon}|J, M\rangle}{E_f-(E_J+\hbar\omega)}\,,
\label{ac}
\end{eqnarray}
where energies $E_{J}$ are the eigenvalues of $H_{\rm rot}$, $\vec{d}_{\rm tr}$, $\vec{\epsilon}$, and $\omega$ are
 the molecular transition electric dipole moment operator,
the laser polarization, and the laser angular frequency, respectively.
The summations over $J$, $M$, $M'$, and $m_k$ only contain basis functions in the low-energy space.
The summation  $f$ in Eq.~\eqref{ac}
is over all ro-vibrational states and continua of excited electronic states with energies $E_f$ excluding their Zeeman, hyperfine, and DC-Stark shifts. We have included  previously studied \cite{Kotochigova:2006, Kotochigova:2010} excited electronic states that dissociate to limits where only one of Rb or Cs is excited to its energetically-lowest excited $n$P state. In this work,  we are interested in the regime where the AC-Stark shift is much smaller than the rotational constant.
Thus, in writing Eq.~\eqref{ac}, couplings between the states with different orbital angular momenta $J$
are neglected. Finally, $\epsilon_0$, $c$, and $\hbar$ are the vacuum permittivity, the speed of light in vacuum, and the reduced Planck's constant, respectively.

We diagonalize   Eq.~(\ref{h_eff}) in the basis $|J, M; m_1, m_2\rangle$ including $J \leq$ 20 to find eigenenergies $E_i$ and corresponding eigenstates $|i\rangle$ of the molecular system. The dynamic polarizability of an eigenstate is  $-\partial E_i/\partial I$.
By mapping out the intensity-dependence of the eigenenergies of the effective low-energy Hamiltonian,
we obtain the dynamic polarizabilities for various rotational states. The electric field, magnetic field,
and laser frequency serve as our tuning parameters which can be manipulated, as shown in the following discussions,
to realize various magic trapping conditions. Although, in this work we focus our discussion on the $^{87}$Rb$^{133}$Cs molecule, the extension to other diatomic alkali molecules is implied.

\section{Zeeman Splittings and DC-Stark Shifts in RbCs molecules}
\label{spectrum}

\begin{figure}
\includegraphics[trim=0.8cm 0cm 0cm 0cm, width=0.85\textwidth]{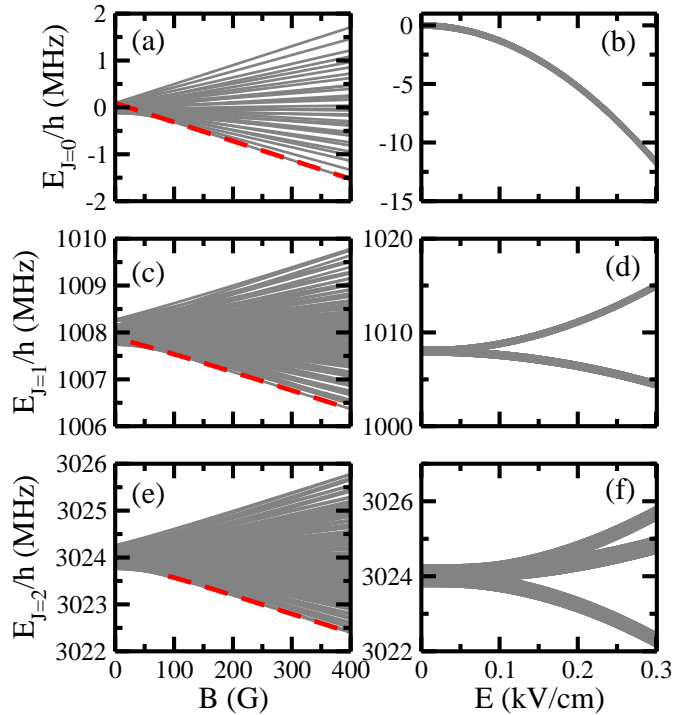}
\vspace{-2cm}
\caption{The hyperfine energy levels for the $J=0$, $1$, and $2$ manifolds
as functions of the magnetic field strength $B$ (the left column (a), (c), and (e) panels, respectively)
and the static electric field strength $E$ applied parallel to a magnetic field of $B=181$\,G (the right column (b), (d), and (f) panels, respectively).
The red dashed lines in (a), (c), and (e) mark the
target trapping state (see text).
Panel (b) consists of a band with 32 energy levels.
Panel (d) consists of two bands;
the upper one contains 32 energy levels with $M=0$
and the lower one 64 energy levels with $M=\pm 1$.
Panel (f) consists of three bands;
the upper one contains 32 energy levels with $M=0$,
the middle one 64 energy levels with $M=\pm 1$,
and the lower one 64 energy levels with $M=\pm 2$.
}
\label{fig1}
\end{figure}

The nuclear spins of $^{87}$Rb and $^{133}$Cs
atoms are $I_1 = 3/2$ and $I_2 =7/2$, respectively.
Because of the multiple
combinations of the atomic nuclear spin projections
and the molecular orbital angular momentum projections,
there exist $(2J+1)(2I_1 +1)(2I_2+1)$ energy levels that are
associated with the rotational state with orbital angular momentum $J$.
In the presence of the magnetic field,
the static electric field, and
the hyperfine interactions,
these ``near" degenerate energy levels split.
Before we discuss the magic trapping conditions,
it is necessary to select
the best target states to be trapped
among these levels
for each rotational state.

The left column of Fig.~\ref{fig1}
shows the magnetic field strength dependence
of the rotational energy manifold $E_J$
$(J=0, 1, 2)$ with vanishing static electric field.
In the weak magnetic field regime,
the splitting between the levels of the same energy manifold
are dominated by the hyperfine interactions.
In this regime,
the total angular momentum
$\vec{F}^2 = (\vec{J} + \vec{I}_1 + \vec{I}_2)^2$
and the total projection
$M_F = M + m_1 + m_2$
are approximately good quantum numbers
which means that
the eigenstates consist of strong admixture of states
with different nuclear spin projections.
The level repulsion is strong in this regime, leading to quadratic Zeeman shifts dominating over linear Zeeman shifts for $B<50$\,G
in the left column of Fig.~\ref{fig1}. With increasing magnetic field strength, the linear Zeeman shift dominates.
Due to the differences in the various $g$-factors, $g_r=0.0062$, $g_1=1.836(3)$, and $g_2=0.738(1)$
in Eq.~\eqref{zeeman} for $^{87}$Rb$^{133}$Cs~\cite{Aldegunde:2008,Gregory:2016},
the projections $M$, $m_1$, and $m_2$ are all approximately good quantum numbers
in the high-field regime. Thus, the eigenstates have significantly reduced admixture.
The levels marked by the red dashed lines in the left column of Fig.~\ref{fig1} correspond to the states containing more than $50\%$ occupation in the $|J, M=0; m_1=3/2, m_2 =7/2\rangle$ component. Note this corresponds to the spin-stretched state in $J=0$ and is the initial state created in experiments~\cite{Takekoshi:2014,Molony:2014}. We select these states as our target trapping states.
For $B>150$\,G, the admixture of the other components into the target trapping states is less than $20\%$
for $J=0, 1$, and $2$.
The red dashed lines in Fig.~\ref{fig1} (c) and Fig.~\ref{fig1} (e)
terminate around $B=25$\,G and $B=100$\,G, respectively,
because no target state can be identified in the small $B$ regime
due to strong admixture.
In this paper,
we focus on a magnetic field strength of $B=181$\,G
according to the experimental work~\cite{Molony:2014}.

To further suppress the admixture of other components into our target trapping states, we take advantage of an applied static electric field. Due to the small magnitude of the rotational $g$-factor $g_r$, the states with different orbital angular momentum projections $M$ and the same nuclear spin projections are close in the spectrum. For example, the energy splitting between states with a unit difference in the orbital angular momentum projection $M$ is $\sim 10$\,kHz for $B=181$\,G.
A static electric field along the magnetic field direction
separates the levels with different absolute values of $M$
in the spectrum.

The right column of Fig.~\ref{fig1} shows the dependence of the rotational energy manifold $E_J$ $(J=1,2,3)$ on the electric field strength in the presence of a parallel magnetic field of $B=181$\,G.
With increasing $E$,
the energies of the $J=0$ manifold decrease quadratically
[see Fig.~\ref{fig1} (b)]
due to the second-order
level repulsion with the states $|J=1, M=0; m_1, m_2\rangle$.
It turns out that the energies of the
states $|J=1, M=0; m_1, m_2\rangle$ are pushed up.
Due to the level repulsion between the states
$|J=1, M=\pm 1; m_1, m_2\rangle$
with
the states
$|J=2, M=\pm 1; m_1, m_2\rangle$,
the states with $M=\pm 1$ in the $J=1$ manifold are pushed down with
increasing $E$.
Thus,
the static electric field separates the $J=1$ rotational
energy manifold into two bands,
the upper one with $M=0$
and the lower one with $M=\pm 1$
[see Fig.~\ref{fig1} (d)].
Similarly,
for the $J=2$ manifold,
a three-band structure is seen
with the upper, the middle, and the lower
one corresponding to $M=0$, $M=\pm 1$, and $M=\pm 2$, respectively.
For $B=181$\,G,
a static electric field of strength $E=0.1\,\text{kV}/\text{cm}$
already makes
the admixture of the states with finite $M$
into the state with $M=0$ negligible.

\section{AC-Stark Shifts Near the Narrow $\mathrm{X}^1\Sigma^+\rightarrow\mathrm{b}^3\Pi_0$ Transitions}
\label{ac_stark}

\begin{figure}
\includegraphics[trim=0cm 0cm 0cm 0cm, clip,width=0.4\textwidth]{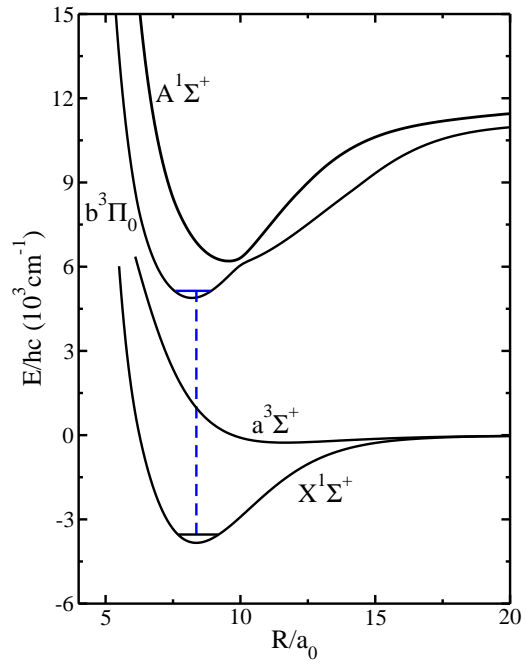}
\caption{
Ground and relevant excited adiabatic relativistic $\Omega=0^+$ potentials of the
$^{87}$Rb$^{133}$Cs molecule as a function of internuclear separation $R$.
The energetically-lowest
potential is identified by non-relativistic label X$^1\Sigma^+$.
The two excited adiabatic potentials
have a narrow avoided crossing at $R_{\rm c}\approx10 a_0$.
For $R<R_{\rm c}$ the electronic
wavefunction of the second adiabat is well described by the non-relativistic b$^3\Pi_0$ symmetry.
For $R>R_{\rm c}$ this state is well described by the A$^1\Sigma^{+}$ symmetry.
The vertical
lines indicate transitions from the $J=0$ trapping state
in the X$^1\Sigma^+$ state to the lowest $J'=1$ ro-vibrational states of the coupled
 A$^1\Sigma^{+}$-b$^3\Pi_0$ complex.
The transition wavelength
is $1146.287$ nm. }
\label{fig2}
\end{figure}

To study the AC-Stark shift of the $^{87}$Rb$^{133}$Cs molecule,
we consider the application of a driving laser field with the angular frequency $\omega$ to induce coupling between the target trapping states
and electronically excited states.
Figure~\ref{fig2} shows the selected relativistic adiabatic $\Omega=0^+$ potential curves
of the $^{87}$Rb$^{133}$Cs molecule,
where $\Omega$ is the total projection quantum number
of the electronic angular momentum and nuclear spins
along the diatomic molecule axis.
The b$^3\Pi_0$ potential and the A$^1\Sigma^+$ potential
are coupled by the spin-orbit coupling terms
which lead to an avoided crossing near $R_c=10a_0$.
Here,
the potentials and the spin-orbit coupling functions
are generated based on the data in {Refs.~\cite{Docenko:2010, Docenko:2011, Mario:2016, Vexiau:2017}.
Due to the spin-orbit coupling,
the few lowest bound states
lying near the bottom of the b$^3\Pi_0$ potential
have some admixture of the A$^1\Sigma^+$ component
which enables
the electric dipole coupling
from these states to
the states of the ground electronic potential X$^1\Sigma^+$.
These transitions are much narrower
than
the transitions to the states with dominant occupation in the A$^1\Sigma^+$ potential.
In this work,
we are particularly interested in the
AC-Stark shift and the dynamic polarizabilities
near these narrow transitions, indicated by the blue dashed line in Fig.~\ref{fig2}.
We denote $\omega_{v'}$
the resonance transition frequency
from the $(v=0, J=0)$ state of the X$^1\Sigma^+$ potential
to the $(v', J=1)$ state of the b$^3\Pi_0$ potential.
For $v'=0$,
the resonance frequency reads $\omega_0 = 2\pi\times 261.533$\,THz
which corresponds to a wavelength of $1146.287$\,nm.
When
the driving laser frequency
$\omega$ is close to the resonance frequency $\omega_{v'}$,
we reference $\omega$
to $\omega_{v'}$ through the detuning $\Delta_{v'} = \omega - \omega_{v'}$.

\begin{figure}
\vspace{1cm}
\includegraphics[trim=0cm 0cm 0cm 3cm, width=0.5\textwidth]{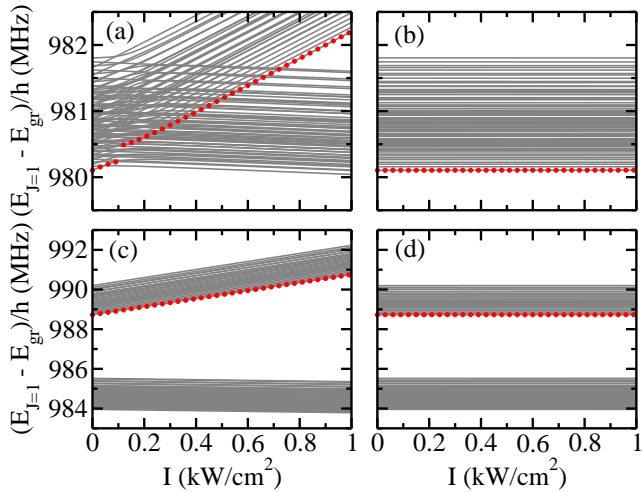}
\caption{
Microwave transition frequencies from the $|J=0, M=0; m_1=3/2, m_2 =7/2\rangle$ ground state to the $J=1$ manifold as a function of the laser intensity for a laser frequency near the resonance transition
to the $v'=0$ vibrational state of the b$^3\Pi_0$ potential.
A magnetic field of strength $B = 181$G is applied in the $z$-direction.
Panels (a) and (c) correspond to a detuning of $\Delta_{v'=0}=2\pi\times 3$\,GHz.
Panels (b) and (d) correspond to a detuning of $\Delta_{v'=0}=2\pi\times 200$\,GHz.
Panels (a) and (b) correspond to vanishing static electric field.
Panels (c) and (d) correspond to a static electric field of $E=0.2\,\text{kV}/\text{cm}$
applied in the $z$-direction.
The red circles in all panels mark the energy level of the target trapping state.
}
\label{fig3}
\end{figure}

Figure~\ref{fig3} shows the impact of the static electric field on the AC-Stark shifts of the microwave transition frequencies from the $|J=0, M=0; m_1=3/2, m_2 =7/2\rangle$ ground state to the $J=1$ rotational energy manifold in the small and large detuning regimes.
The driving laser is linearly polarized
with a polarization parallel to the magnetic field.
The red circles correspond to the target trapping state as discussed in Sec.~\ref{spectrum}.
For the case with the detuning of $\Delta_{v'=0} = 2\pi\times 3$\,GHz
and vanishing static electric fields
[Fig.~\ref{fig3} (a)],
the AC-Stark shifts can be characterized into two bands;
one going up with increasing laser intensity
while
the other staying almost independent of the laser intensity.
The former corresponds to states with $M=0$
while the latter to states with $M=\pm 1$.
As shown by the red circles in Fig.~\ref{fig3} (a),
the energy level of the target trapping state in the $J=1$ manifold
crosses those of the other levels with increasing laser intensity.
These crossings lead to strong level interactions
[see the gap in the red circles
near $I=0.1\text{kW}/\text{cm}^2$ in Fig.~\ref{fig3} (a)],
hence to large hyper-polarizabilities
which makes the system unstable
with respect to fluctuations of the trapping laser intensity.

The level-crossing behavior in the AC-Stark shift
can be avoided by separating the
$M=0$ band and the $M=\pm 1$ band
using a static electric field
as discussed in Sec.~\ref{spectrum}.
Figure~\ref{fig3} (c) shows the AC-Stark shifts
in the presence of a static electric field
of $E=0.2\,\text{kV}/\text{cm}$.
Compared to Fig.~\ref{fig3} (a),
the $M=0$ band lies roughly $5$\,MHz above the $M=\pm 1$ band
for $I=0$.
With increasing laser intensity,
the energy gap between the
$M=0$ band and the $M=\pm 1$ band
keeps increasing.
The energy of the target trapping state does not cross any
of the $M=\pm 1$ states any more.

The level crossings seen in Fig.~\ref{fig3}\,(a) result from the fact that the AC-Stark shift of the target trapping state
is larger than the energy splitting
between the nearest neighbor hyperfine levels.
With larger laser detuning,
the differential AC-Stark shift is greatly reduced.
For example,
for a detuning of $\Delta_{v'=0} = 2\pi \times 200$\,GHz
as shown in Fig.~\ref{fig3}\,(b),
the level crossings between the target trapping state
and the other states in the $J=1$ manifold
disappear for the laser intensity regime shown here.
A finite static electric field still separates the
$M=0$ band from the $M=\pm 1$ band as shown in Fig.~\ref{fig3}\,(d),
which does make the system more robust,
but is not necessary in this case.

In the following discussion of dynamic polarizabilities,
we describe the detuning as near-resonance when $\Delta_{v'}<2\pi\times 10$\,GHz and as medium-detuned otherwise.
According to the above discussion,
the static electric field is always turned on
for the near-resonance cases
and not mandatory for the far-detuned cases.
This setup makes
our results independent of the laser intensity
in a broad intensity regime for both cases.

\section{Magic conditions for multiple rotational states}
\label{magic}

We may identify magic trapping frequencies by searching for crossings among the frequency-dependent dynamic polarizability curves of different rotational states.
We start the discussion with the dynamic polarizabilities $\alpha_J$
near the resonance from which we extract the parallel and perpendicular background polarizabilities
$\alpha_{\text{bg}, \parallel}$ and $\alpha_{\text{bg}, \perp}$ and
the transition width $\Gamma_{0,v'}$.
Given the values of
$\alpha_{\text{bg}, \parallel}$, $\alpha_{\text{bg}, \perp}$, and $\Gamma_{0,v'}$,
it is proved analytically
and verified by our numerical calculations that
there exists a ``near" magic frequency window
for multiple rotational states in the medium-detuned regime between vibrational poles.
By tuning the static electric field,
a true triple magic frequency is found
for the $J=0$, $J=1$, and $J=2$ target trapping states
for the $^{87}$Rb$^{133}$Cs molecule.

\subsection{Near-Resonance Dynamic Polarizabilities}
\label{near_resonance}

\begin{figure}
\includegraphics[trim=0cm 0cm 0cm 0cm, width=0.5\textwidth]{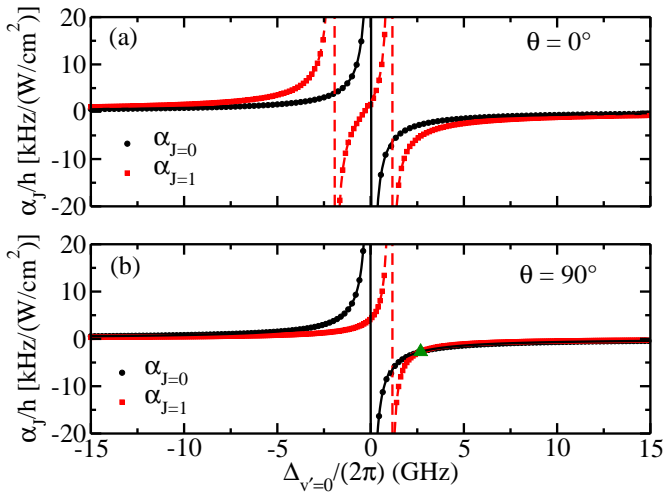}
\caption{
The dynamic polarizabilities near the resonance transition
to the $v=0$ vibrational state of the b$^3\Pi_0$
potential.
A magnetic field of strength $B=181$\,G
and a static electric field of strength $E=0.2\,\text{kV}/\text{cm}$
are applied in the $z$-direction.
The driving laser polarization is (a) parallel and (b) perpendicular to the external static fields.
The black circles and red squares correspond to the numerical results
of the dynamic polarizabilities of the $J=0$ and $J=1$ target trapping state.
The black solid lines and the red solid lines correspond to the
analytical results generated using the Eqs.~\eqref{approx_near_res_J0}
and~\eqref{approx_near_res_J1}.
The green upper triangle in Panel (b)
marks the crossing between the black circles and red squares.
}
\label{fig4}
\end{figure}

In the near-resonance regime, we fix the strength of the static electric field to be $E=0.2\,\text{kV}/\text{cm}$.
The angle between the laser polarization and the magnetic field is denoted $\theta$.
In this case,
the dynamic polarizabilities $\alpha_{J=0}$ of the $J=0, M=0$ target trapping state
and $\alpha_{J=1}$
of the $J=1, M=0$ target trapping state
can be approximated using \cite{Neyenhuis:2012} by
\begin{eqnarray}
\label{approx_near_res_J0}
\alpha_{J=0} =  -\frac{3\pi c^2}{2\omega_{v'}^3}\frac{\Gamma_{0,v'}}{3\Delta_{v'}} + \frac{1}{3}\alpha_{\text{bg},\parallel} + \frac{2}{3}\alpha_{\text{bg}, \perp},
\end{eqnarray}
and
\begin{align}
\label{approx_near_res_J1}
\alpha_{J=1} & =
-\frac{3\pi c^2}{2\omega_{v'}^3}\Big[
\frac{\cos^2(\theta)}{3}
\frac{\Gamma_{0,v'}}{\Delta_{v'}+2B_{v}+2B_{v'}}+\\\nonumber
& \frac{3+\cos^2(\theta)}{15}\frac{\Gamma_{0,v'}}{\Delta_{v'}+2B_{v}-4B_{v'}}\Big]+ \\\nonumber
& \frac{2\cos^2(\theta)+1}{5}\alpha_{\text{bg},\parallel} + \frac{4-2\cos^2(\theta)}{5}\alpha_{\text{bg}, \perp},
\end{align}
respectively.
Here,
the parameters $B_{v}$ and $B_{v'}$
correspond to the rotational constants for the
$v=0$ vibrational state of the X$^1\Sigma^+$ potential
and the $v'=0$ vibrational state of the b$^3\Pi_0$ potential.
The transition width $\Gamma_{0,v'}$
can be calculated via
\begin{eqnarray}
\label{transition_width}
\Gamma_{0,v'} = \frac{\omega_{v'}^3}{3\pi\epsilon_0 \hbar c^3} |\mu_{0,v'}|^2
\end{eqnarray}
where the $\mu_{0,v'}$
is the transition dipole momentum
between the $v=0$ vibrational state of the
X$^1\Sigma^+$ potential
and the $v'$ vibrational state of the
b$^3\Pi_0$ potential.
The parallel and perpendicular
background polarizabilities
$\alpha_{\text{bg}, \parallel}$
and $\alpha_{\text{bg}, \perp}$
contain
the contributions from all the
far-detuned rovibronic states
with $\Omega = 0$ and $\Omega = 1$, respectively~\cite{Kotochigova:2006,Vexiau:2017,Li:2017}.
For $^{87}$Rb$^{133}$Cs, we find 
$B_{v}=2\pi\times 0.490$\,GHz,
$B_{v'}=2\pi\times 0.510$\,GHz,
$\Gamma_{0,v'=0}=2\pi\times 15.5\,\text{kHz}$,
$\alpha_{\text{bg},\parallel}=h\times 0.127$\,kHz/(W/cm$^2$),
and
$\alpha_{\text{bg},\perp}=h\times 0.0340$\,kHz/(W/cm$^2$).
Experimentally, these values can be extracted by
fitting the measured dynamic polarizability curves
near the poles.

Figure~\ref{fig4} shows the dynamic polarizabilities
for laser polarizations parallel and perpendicular to the magnetic field direction in the near-resonance regime.
The symbols correspond to
the numerical results and the lines show the analytical results
generated using Eqs.~\eqref{approx_near_res_J0} and~\eqref{approx_near_res_J1}.
The agreement in both cases is excellent.
As can be seen, there is no crossing between the $\alpha_{J=0}$ curve
and the $\alpha_{J=1}$ curve in the near-resonance regime for $\theta=0^{\circ}$.
According to Eq.~\eqref{approx_near_res_J1},
the dynamic polarizability $\alpha_{J=1}$
can be tuned by varying the polarization direction of the driving laser.
For example,
for $\theta=90^\circ$,
the term
in the first row of Eq.~\eqref{approx_near_res_J1} inside the square bracket
vanishes
and the pole structure at $\Delta_{v'=0} = -2\pi\times2.00$~GHz is missing, as shown by the red squares in Fig.~\ref{fig4} (b).
In addition,
the pole at $\Delta_{v'=0} = 2\pi\times 1.06$~GHz is slightly
narrower compared to
the $\theta =0^\circ$ case.
In this case, the $\alpha_{J=1}$ curve crosses the $\alpha_{J=0}$
curve at the magic detuning of $2\pi\times 2.68$~GHz, as shown by the green upper triangle in Fig.~\ref{fig4} (b).
The value of the polarizability at the magic detuning is $-h\times 2.71$\,kHz/(W/cm$^2$).
The negative polarizability indicates that the molecules can be trapped at the nodal point of an optical lattice
where the laser intensity is the local minimum.
This trapping condition is beneficial for also minimizing heating and loss from incoherent photon scattering.

\subsection{Multiple Magic Frequency Window}
\label{triplet_magic}

For arbitrary $J$,
we derive
the general formula for
the dynamic polarizability near the resonance transition to
one of the states
of the b$^3\Pi_0$ potential,
\begin{align}
\label{general}
\alpha_J &= -\frac{3\pi c^2}{2\omega_{v'}^3}
\left[A_J(\theta) \frac{\Gamma_{0,v'}}{\Delta_{v'} +  L_{J}}
+ B_J(\theta) \frac{\Gamma_{0,v'}}{\Delta_{v'} + R_{J}}\right]\\\nonumber
&+ \left[A_J(\theta) + B_J(\theta)\right]\alpha_{\text{bg}, \parallel}
+ \left[1- A_J(\theta) - B_J(\theta)\right]
\alpha_{\text{bg}, \perp},
\end{align}
where the pole positions $L_J$ of the left branch
and $R_J$ of the right branch read
\begin{eqnarray}
L_{J}= J(J+1) B_v - [J(J-1)-2]B_{v'},
\end{eqnarray}
and
\begin{eqnarray}
R_{J}=J(J+1) B_v - [(J+1)(J+2)-2]B_{v'},
\end{eqnarray}
respectively.
The angular factors $A_J(\theta)$
and $B_J(\theta)$ in Eq.~\eqref{general} are,
\begin{eqnarray}
\label{factor_A}
A_J(\theta) =
\left\{
\begin{aligned}
&\frac{(J+1)(J-1)}{2(2J+1)(2J-1)} + &\\
&\frac{J^2+1}{2(2J+1)(2J-1)}\cos^2(\theta) & J>0 \\
 &0 & J=0,
\end{aligned}
\right.
\end{eqnarray}
and
\begin{align}
\label{factor_B}
B_J(\theta) = &\frac{(J+2)(J+1)}{2(2J+3)(2J+1)} +&\\\nonumber
&\frac{J(J+1)}{2(2J+3)(2J+1)} \cos^2(\theta).
\end{align}
By Taylor-expanding the right hand side of
Eq.~\eqref{general}
with respect to $L_J$ and $R_J$,
we obtain,
\begin{align}
\label{general_app}
\alpha_J &=
\left[A_J(\theta) + B_J(\theta)\right]
\left(-\frac{3\pi c^2}{2\omega_{v'}^3}\frac{\Gamma_{0,v'}}{\Delta_{v'}}
+\alpha_{\text{bg}, \parallel} - \alpha_{\text{bg}, \perp}\right)+\\\nonumber
&\alpha_{\text{bg}, \perp} + T_J(\Delta_{v'}, \theta),
\end{align}
where the remaining term $T_J(\Delta_{v'},\theta)$ reads,
\begin{align}
\label{remaining}
T_J(\Delta_{v'}, \theta)=&\frac{3\pi c^2}{2\omega_{v'}^3}\frac{\Gamma_{0,v'}}{\Delta_{v'}^2}
\left[A_J(\theta)L_J+B_J(\theta)R_J\right]+\\\nonumber
&\mathcal{O}\left(\frac{\Gamma_{0,v'}L_J^2}{\Delta_{v'}^3}\right)
+\mathcal{O}\left(\frac{\Gamma_{0,v'}R_J^2}{\Delta_{v'}^3}\right).
\end{align}
Based on Eq.~\eqref{general_app},
we can always find a detuning $\Delta_{v',\text{cr}}$ such that,
\begin{eqnarray}
\label{critical_pola}
\alpha_J = \alpha_{\text{bg},\perp} + T_J(\Delta_{v',\text{cr}},\theta),
\end{eqnarray}
where,
\begin{eqnarray}
\label{critical_detuning}
\Delta_{v',\text{cr}} = \frac{3\pi c^2}{2\omega_{v'}^3}\frac{\Gamma_{0,v'}}{\alpha_{\text{bg},\parallel}-\alpha_{\text{bg},\perp}}.
\end{eqnarray}
For the transitions
with $\Delta_{v',\text{cr}}$ lying in the medium-detuned regime,
i.e.,  $|\Delta_{v',\text{cr}}|\gg \left|L_J\right|$, $|\Delta_{v',\text{cr}}|\gg \left|R_J\right|$,
and $|\Delta_{v',\text{cr}}|\gg \Gamma_{0,v'}$,
the remaining term $T_J(\Delta_{v',\text{cr}},\theta)$ can be neglected.
In this case,
both the $\theta$-dependence and the $J$-dependence of $\alpha_J$
in Eq.~\eqref{critical_pola} disappear, indicating that the frequency-dependent
dynamic polarizabilities of all rotational states pass through the same fixed point; the trap is magic for all rotational states at this laser detuning. The multiple
magic frequency is approximately given by Eq.~\eqref{critical_detuning}
and the value of the dynamic polarizability is approximately equal to
the background perpendicular dynamic polarizability $\alpha_{\text{bg},\perp}$.

\begin{figure}
\vspace*{0.5cm}
\includegraphics[trim=0.5cm 0cm 0cm 0cm, width=0.5\textwidth]{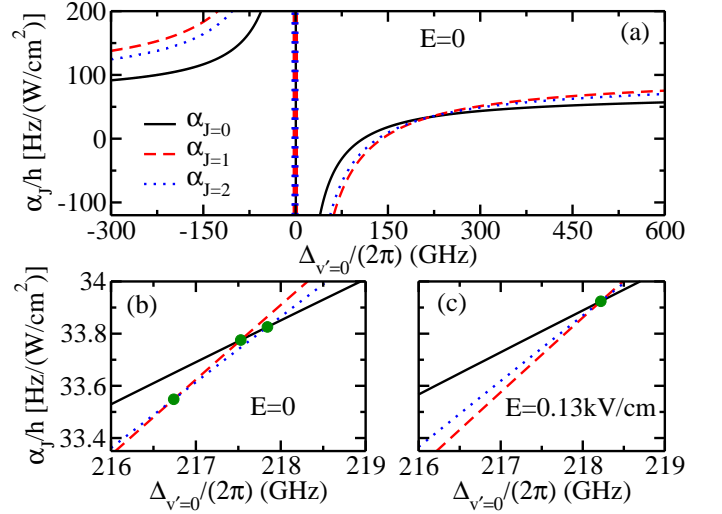}
\caption{
The triple magic conditions for $J=0$, $1$, and $2$ rotational states
near the resonance transition to the $v=0$ state of the b$^3\Pi_0$ potential.
A magnetic field $B=181$G is applied in the $z$-direction.
The laser polarization is parallel to the magnetic field.
The circles mark the
crossings between different curves in (b) and (c).
The static electric field is vanishing in (a) and (b).
A finite static electric field of $E=0.13\,\text{kV}/\text{cm}$ is applied along the $z$-direction in (c).
A near triple magic condition exists in (b) and a true triple magic condition exists in (c).
}
\label{fig5}
\end{figure}

Figure~\ref{fig5}\,(a) shows the triple crossing magic frequency
for $\alpha_{J}$ with $J=0$, $1$, and $2$ near the resonance transition
to the $v'=0$ vibrational states of the b$^3\Pi_0$ potential.
The three curves cross each other
in the detuning window of
$2\pi\times216$\,GHz to $2\pi\times219$\,GHz, as highlighted in Fig.~\ref{fig5}\,(b).
Evaluating Eq.~\eqref{critical_detuning} using the values of the transition width and the background polarizabilities
obtained in Sec.~\ref{near_resonance}, the predicted magic frequency corresponds to a detuning of $2\pi\times 240$\,GHz.
The difference comes from the higher order corrections
in the remaining term $T_J(\Delta_{v'},\theta)$.
The range of the $\alpha_J$ values in Fig.~\ref{fig5}\,(b) is
consistent with the value of $\alpha_{\text{bg},\perp}$
as calculated in Sec.~\ref{near_resonance}.
Even though the three curves do not intersect
each other at the same frequency,
their values are very close in the frequency window
shown in Fig.~\ref{fig5}\,(b).
The percent difference $\left|\alpha_J - \alpha_{J'}\right|/|\alpha_{J'}|$
for any pair of $J$ and $J'$ in Fig.~\ref{fig5}\,(b) is less than $0.6\%$
within the detuning range of $2\pi\times 3$\,GHz,
which makes the magic trapping condition robust to
uncertainty in the trapping laser frequency.
This near triple magic frequency window
can be tuned to a true triple magic frequency
by adding a weak static electric field.
Figure~\ref{fig5}\,(c) shows that the three
curves cross at $\Delta_{v'=0} = 2\pi\times 218.22$\,GHz
for $E = 0.13$\,kV/cm. The value of the polarizability at this detuning is 
$\alpha_J=h\times0.03392$\,kHz/(W/cm$^2$).

\begin{figure}
\includegraphics[trim=0cm 0cm 0cm 0cm, width=0.5\textwidth]{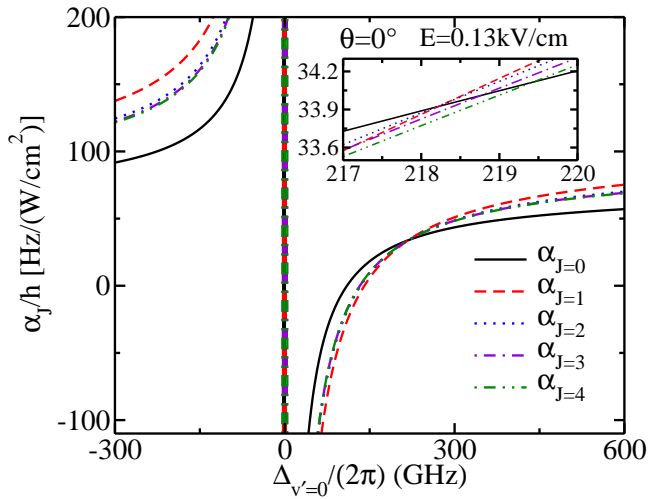}
\caption{
The dynamic polarizabilities near the resonance transition
to the $v=0$ vibrational state of the b$^3\Pi_0$
potential for multiple rotational states up to $J=4$.
A magnetic field of $B=181$\,G and a static electric field of $E=0.13\,\text{kV}/\text{cm}$
are applied in the $z$-direction.
The laser polarization is parallel to the $z$-axis.
The insets show the zoom-in of the ``near magic'' frequency window
in which the polarizabilities of many rotational state are
either crossing or close to each other.
}
\label{fig6}
\end{figure}

Our theory also predicts that
the triple magic frequency window also holds for higher rotational states.
Figure~\ref{fig6} shows the $\alpha_J$ curves up to $J=4$
for the parallel driving case
in the presence of the static electric field of strength $E=0.13$~kV/cm.
It can seen that all the values of $\alpha_J$ are very close to $\alpha_{\text{bg},\parallel}$
in the same magic frequency window as discussed before.
A further zoom-in of the magic frequency window, shown in the inset of Fig.~\ref{fig6}, indicates that $\alpha_{J=3}$ and $\alpha_{J=4}$
almost run parallel to $\alpha_{J=2}$ and, consequently, do not pass through the triple magic frequency point
for the $\alpha_{J=0,1,2}$ curves.
The higher rotational states make the contribution from the remaining term
$T_{J}(\Delta_{v',\text{cr}},\theta)$ more important
due to larger values of $|L_J|$ and $|R_J|$.
Thus, no crossings among the polarizability curves of higher $J$ values
are expected within the magic frequency window.

The similarity of
the $\alpha_J$ curves in the medium-detuned regime
with increasing $J$ values
is explained by the asymptotic behavior of the angular factors $A_J(\theta)$ and $B_J(\theta)$
in Eqs.~\eqref{factor_A} and~\eqref{factor_B} in the large $J$ limit.
Expanding $A_J(\theta)$ and $B_J(\theta)$ in terms of $1/J$,
we obtain,
\begin{eqnarray}
\label{factor_A_limit}
A_{J}(\theta) = \frac{1+\cos^2(\theta)}{8}
+\mathcal{O}\left(\frac{1}{J^2}\right),
\end{eqnarray}
and,
\begin{eqnarray}
\label{factor_A_limit}
B_{J}(\theta) = \frac{1+\cos^2(\theta)}{8} +
\frac{\sin^2(\theta)}{8J}
+\mathcal{O}\left(\frac{1}{J^2}\right).
\end{eqnarray}
With increasing $J$,
the leading order terms of
both $A_J(\theta)$ and $B_J(\theta)$
are independent of the value of $J$;
hence the expression for $\alpha_J$
in Eq.~\eqref{general_app} becomes the same for all $J$, neglecting the remaining $T_{J}(\Delta_{v'},\theta)$ term.
Thus,
for large $J$,
the various $\alpha_J$ curves are close and almost parallel to each other in the medium-detuned regime.
Combining the true triple magic condition for the lower
$J$ values and the similarity between $\alpha_J$ for higher $J$ values, leads to a ``near magic'' trapping window for multiple rotational states that should be possible to realize experimentally.

\begin{figure}
\hspace*{0.0cm}
\includegraphics[width=0.5\textwidth]{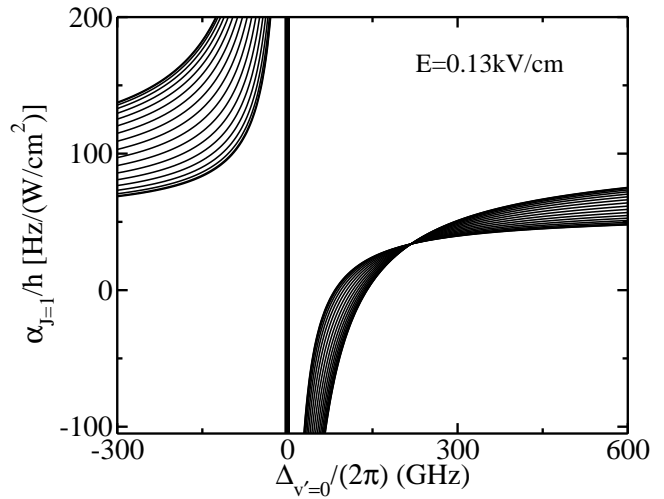}
\caption{
The dynamic polarizabilities $\alpha_{J=1}$
near the resonance transition
to the $v'=0$ vibrational state of the b$^3\Pi_0$
potential for various driving laser polarization directions.
The angle $\theta$ is scanned from $0^\circ$ to $90^\circ$
in $5^\circ$ increments.
A magnetic field of $B=181$\,G and a static electric field of $E=0.13$\,kV/cm
are applied in the $z$-direction.
}
\label{fig7}
\end{figure}

The $\theta$-independence of $\alpha_J$
within the multiple magic frequency window
is also verified by our numerical results.
Figure~\ref{fig7} shows
the dynamic polarizability $\alpha_{J=1}$ for angles
between $0^\circ$ and $90^\circ$.
All the curves nearly cross the same point around the detuning of $2\pi\times 218$\,GHz.

Based on all the results and observations discussed above,
we conclude that the existence of the multiple magic frequency window presents
a frequency region of a few gigahertz within which the system is super robust
with respect to the fluctuations of the trapping laser frequency
and the polarization direction for arbitrary rotational states. Within this window long-rotational coherences should be possible on multiple rotational transitions in the $^{87}$Rb$^{133}$Cs molecule.

\subsection{Criteria for the Multiple Magic Frequency Window}
\label{criteria}

\begin{figure}
\hspace*{0.3cm}
\includegraphics[trim=0cm 0cm 0cm 0cm, width=0.9\textwidth]{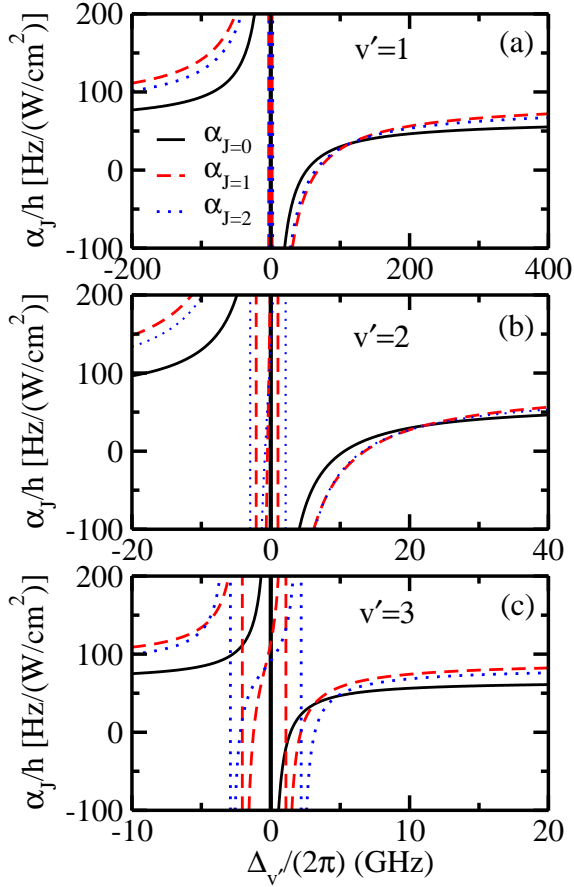}
\caption{
The dynamic polarizabilities of the $J=0$, $1$, and $2$
rotational states near the resonance transitions
to the (a) $v'=1$, (b) $v'=2$, and (c) $v'=3$ vibrational states of the b$^3\Pi_0$
potential.
A magnetic field of $B=181$\,G is applied in the $z$-direction.
No static electric field is applied.
The black solid, red dashed, and blue dotted lines
correspond to the dynamic polarizabilities of $J=0$, $1$, and $2$
rotational states, respectively.
A near triple magic condition exists in (a) and (b)
but not in (c).
}
\label{fig8}
\end{figure}

The existence of the multiple magic frequency window
relies on the condition that the remaining $T_{J}(\Delta_{v'},\theta)$ term in Eq.~\eqref{critical_pola}
is much smaller than the $\alpha_{\text{bg},\perp}$ and thus can be neglected.
Taking the leading order term of $T_J(\Delta_v, \theta)$
in Eq.~\eqref{remaining},
the condition $|T_J(\Delta_{v', \text{cr}}, \theta)|\ll |\alpha_{\text{bg},\perp}|$
yields a lower bound on the transition width $\Gamma_{0,v'}$
in terms of the background polarizabilities and rotational constants,
\begin{align}
\label{conditions}
\Gamma_{0,v'} \gg
\frac{2\omega_{v'}^3}{3\pi c^2}
\frac{\left(\alpha_{\text{bg},\parallel}-\alpha_{\text{bg},\perp}\right)^2}{|\alpha_{\text{bg},\perp}|}
\sqrt{B_v^2 +B_{v'}^2}.
\end{align}
For $^{87}$Rb$^{133}$Cs molecules
near the narrow transitions
to the bottom of the b$^3\Pi_0$ potential,
the right hand side of Eq.~\ref{conditions} is equal to $2\pi\times 0.125$~kHz.
As the transition linewidth $\Gamma_{0,v'}$ decreases with increasing $v'$, this condition puts a constraint
on the number of vibrational poles
around which the multiple magic frequency window exists.

Figure~\ref{fig8} shows $\alpha_J$ for $J=1$, $2$, and $3$
near the $v'=1$, $2$, and $3$ vibrational poles
at the bottom of b$^3\Pi_0$ potential.
With increasing $v'$, the transition is narrower
and the triple crossing moves towards
the pole of $\alpha_J$.
The transition widths are $\Gamma_{0,v'=1}=2\pi\times 6.84$\,kHz for the $v'=1$ pole and $\Gamma_{0,v'=2}=2\pi\times 1.44$\,kHz for the $v'=2$ pole. Triple crossings can be seen around
$\Delta_{v'=1} = 2\pi\times 120$\,GHz for the $v'=1$ vibrational pole (Fig.~\ref{fig8}\,(a))
and around $\Delta_{v'=2} = 2\pi\times 22$\,GHz near the $v'=2$ vibrational pole (Fig.~\ref{fig8}\,(b)).
For $v'=3$,
the transition width $\Gamma_{0,v'=3}$ is
$2\pi\times 0.206$~kHz
which is already close to the lower bound.
Thus,
no triple crossings can be seen in Fig.~\ref{fig8} (c).

\subsection{Imaginary Polarizability in the Magic Trapping Window}
\label{imaginary}

\begin{figure}
\includegraphics[trim=0cm 0cm 0cm 1cm, width=0.5\textwidth]{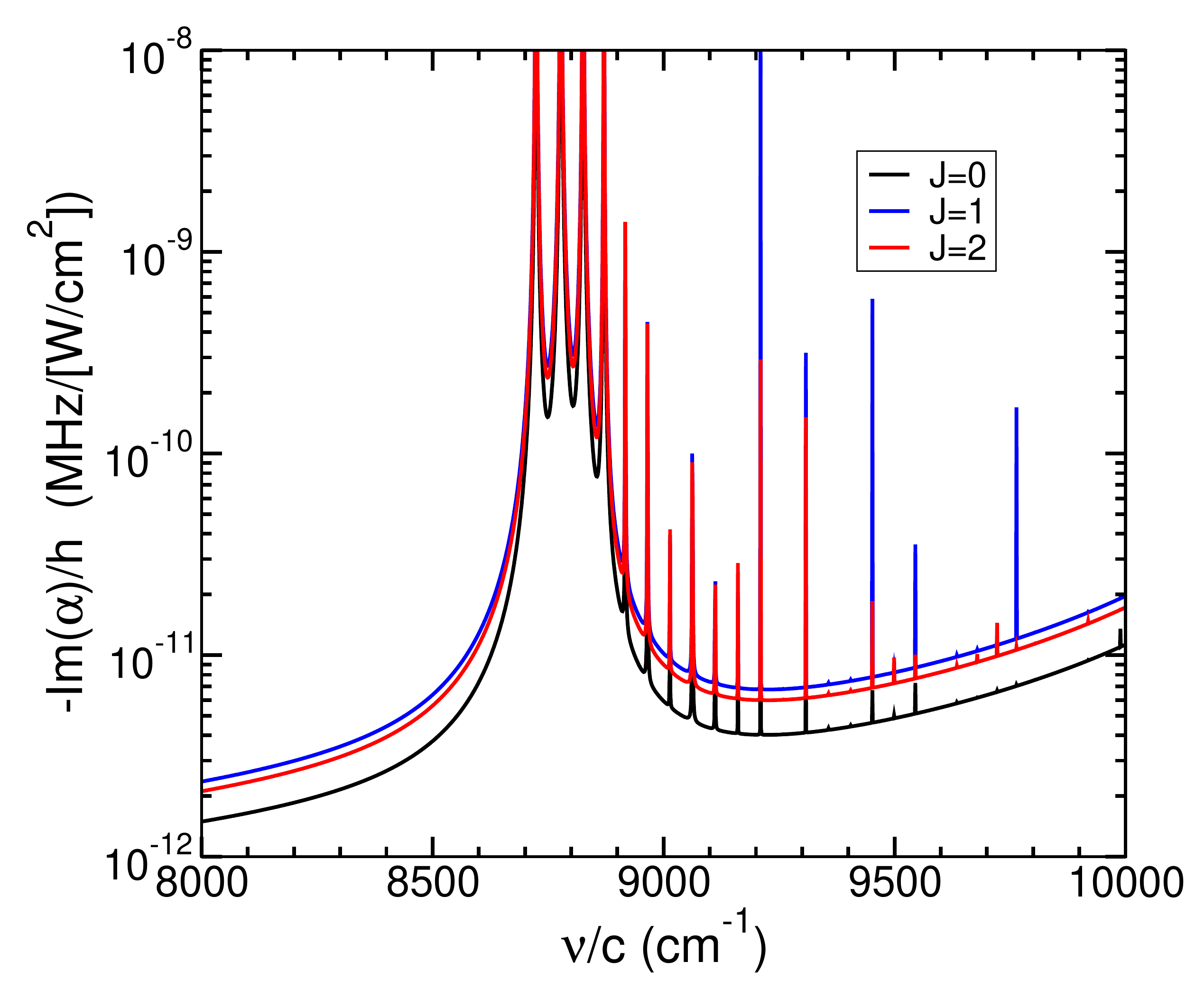}
\caption{The imaginary polarizabilities for the $v=0$, $J=0$, $J=1$ and $J=2$, $M=0$ states of the X$^1\Sigma^+$ potential near the resonance transitions to the lower vibrational states of the b$^3\Pi_0$ potential for $\sigma_{z}$ polarization of the trapping light.
}
\label{imag_pola}
\end{figure}

Light-induced decoherence of rovibrational levels of a polar molecule is often characterized by the imaginary part of the polarizability \cite{Chotia:2012}, which accounts for  losses due to spontaneous emission and  other decay mechanism of intermediate electronically excited states. Here, we evaluate the imaginary part of the complex molecular dynamic polarizability $\alpha(\hbar\omega,\vec{\epsilon})$ as
\begin{eqnarray}
   \lefteqn{ \alpha(\hbar\omega,\vec{\epsilon}) = }
      \label{eqpolar}
\\
  &&\frac{1}{\epsilon_0c}
   \sum_{f} \frac{(E_f - ih\gamma_f/2 - E_i)}{(E_f - ih\gamma_f/2 - E_i)^2 - (\hbar\omega)^2}
     \times |\langle f|{\vec d_{tr}} \cdot \vec{\epsilon}|i\rangle|^2\,, \nonumber
\end{eqnarray}
assuming that each of these intermediate $E_f$ state has a line width $\gamma_f$ equal to 6 MHz, the atomic line width of Rb 5p($^2$P) state. This assumption is justified by  previous calculations of the imaginary polarizability of rovibrational levels of ground state KRb  molecules \cite{APetrov:2013} and a comparison of $\alpha_{\text{imag}}$ with an experimentally measured value \cite{Chotia:2012}.
The sum over $f$ in Eq.~\ref{eqpolar} is limited to transitions to relativistic electronic excited potentials that dissociate to either a singly excited Rb or a singly excited Cs atom.

Figure~\ref{imag_pola} shows the calculated imaginary part of the  polarizability of the $v=0,J=0,1,2$ X$^1\Sigma^+$
states as functions of laser frequency.
By construction the imaginary part is negative. It is several orders of magnitude smaller than the real part.
The resonances in the graph correspond to poles due to the lowest vibrational $v^{\prime}$  of the $\Omega=0$ relativistic component of the b$^3\Pi_0$ potential. 
For a detuning of $\Delta_{v'=0} = 2\pi\times 218$\,GHz close to the triple magic frequency shown in Fig.~\ref{fig5}, the value of the imaginary part of the polarizability is $1.0\times 10^{-9}$\,kHz/(W/cm$^2$). For comparison, the polarizability at this detuning is $\alpha_J=h\times0.03392$\,kHz/(W/cm$^2$), as stated earlier.

\section{Discussion}
\label{discussion}

Although all the results above
are derived by considering transitions to the b$^3\Pi_0$ potential,
similar results to Eqs.~\eqref{critical_pola}
and~\eqref{critical_detuning} are found
for $\Omega = 1$ potentials
with $\alpha_{\text{bg},\perp}$ replaced by $\alpha_{\text{bg},\parallel}$
and vice versa.
These observations indicate that
any rovibrational pole that is associated
with a resonance transition to the state
with quantum number $\Omega$
can be used to cancel
the contributions to the rank-2 dynamic polarizability tensor
from all the other far-detuned states
with the same quantum number $\Omega$.
What remains is the contribution to the dynamic polarizability
from the states with different $\Omega$.
This cancellation happens at a frequency
that is independent of the rotational quantum number $J$
and the polarization direction of the laser.

Even though
the derivation of the equations in
Sec.~\ref{triplet_magic}
is ``universal", i.e.
independent of the molecule species, the existence of the magic frequency window
does require certain conditions to be fulfilled.
For example, Eq.~\eqref{conditions}
gives us a lower bound on the transition width.
For heavier molecules,
such as $^{87}$Rb$^{133}$Cs,
this condition can be satisfied near
the narrow transitions to the bottom of b$^3\Pi_0$ potential,
since the spin-orbit coupling effect is stronger
and the rotational constants, $B_{v}$ and $B_{v'}$, are smaller.
For $^{23}$Na$^{87}$Rb,
we also find that
the multiple magic frequency window exists near the narrow transitions to the b$^3\Pi_0$ potential.
However, compared to $^{87}$Rb$^{133}$Cs, the window only exists near the $v'=0$ and $v'=1$ vibrational poles
and missing near the $v'=2$ pole.

Here, we emphasise that the condition on the lower bound of the transition width given by Eq.~\eqref{conditions}
is not the only criteria for the existence of
the multiple magic frequency window.
Eq.~\eqref{conditions} allows the multiple magic frequency window to also be found near to broad transitions. However, in this case, the predicted magic frequency position
in Eq.~\eqref{critical_detuning}
cannot be larger than the energy spacing between
two nearest neighbor vibrational poles (i.e., $|\Delta_{v', cr}|\ll\left|\omega_{v'\pm1}-\omega_{v'}\right|$).
This condition puts
an upper bound for the transition width,
\begin{eqnarray}
\label{upper}
\Gamma_{0,v'}\ll \frac{2\omega_{v'}^3}{3\pi c^2}
\left|\alpha_{\text{bg},\parallel}-\alpha_{\text{bg},\perp}\right|\times
\left|\omega_{v'\pm1}-\omega_{v'}\right|,
\end{eqnarray}
where
the ``$+/-$" should be used for the positive/negative value of
$\alpha_{\text{bg},\parallel}-\alpha_{\text{bg},\perp}$.
This condition is very easily satisfied near the narrow transitions,
however, it needs to be examined near to the broad ones.
This condition implies that we need to be in the ``medium-detuned" regime
to find the multiple magic frequency window.

Although the existence of the multiple magic frequency windows needs to be checked case-by-case,
the results derived in this work will greatly benefit the search for them.
In experiments,
the background values
of the polarizabilities and the transition widths
can both be straightforwardly measured.
According to Eq.~\eqref{critical_detuning},
the magic detuning can then be predicted based entirely upon these measured values.

\section{Conclusion}
\label{conclusion}

We have investigated magic-wavelength trapping of ultracold bialkali molecules in the vicinity of weak optical transitions from the vibrational ground state of the X$^1\Sigma^+$ potential to low-lying rovibrational states of the b$^3\Pi_0$ potential, focussing our discussion on the $^{87}$Rb$^{133}$Cs molecule. We have shown that a magic trapping frequency window for multiple rotational states exists between two nearest neighbor vibrational poles, far away from any rotational poles. Within this window, the laser trapping is ``near magic" for multiple rotational states simultaneously and is exactly magic for pairs of neighboring rotational states at specific laser frequencies. Moreover, the ``near magic" frequency window can be tuned to a true magic frequency for the lowest three rotational states by applying an experimentally accessible DC electric field. This true triple magic condition is expected to be useful for future studies of synthetic spin-1 systems using ultracold molecules.

We have derived a set of criteria that must be fulfilled to ensure the existence of such magic frequency windows and have also presented an analytic expression for the position of the frequency window in terms of a set of experimentally measurable parameters. These will provide a straightforward, self-consistent approach to search for the magic trapping frequency window in future experiments. We expect the realization of optical traps which are simultaneously magic for multiple rotational states will enable the implementation of highly tunable models in quantum magnetism~\cite{Gorshkov:2011b} and the mapping of many rotational levels onto a synthetic dimension~\cite{Sundar:2018}. More broadly, our work is relevant in settings where there is a need to control the relative polarizabilities of different molecular rotational states, facilitating, for example, the study of Hopf insulators in dipolar systems~\cite{Schuster:2019}.

\section{Acknowledgements}
\label{acknowledgement}

SLC acknowledges support from the UK Engineering and Physical Sciences Research Council (grant numbers EP/P01058X/1 and EP/P008275/1). Work at Temple University is supported by the Army Research Office Grant No. W911NF- 17-1-0563, the U.S. Air Force Office of Scientific Research Grant No. FA9550-19-1-0272 and the National Science Foundation Grant No. PHY-1908634.

\appendix

\bibliography{magic}

\begin{thebibliography}{82}
\expandafter\ifx\csname natexlab\endcsname\relax\def\natexlab#1{#1}\fi
\expandafter\ifx\csname bibnamefont\endcsname\relax
  \def\bibnamefont#1{#1}\fi
\expandafter\ifx\csname bibfnamefont\endcsname\relax
  \def\bibfnamefont#1{#1}\fi
\expandafter\ifx\csname citenamefont\endcsname\relax
  \def\citenamefont#1{#1}\fi
\expandafter\ifx\csname url\endcsname\relax
  \def\url#1{\texttt{#1}}\fi
\expandafter\ifx\csname urlprefix\endcsname\relax\def\urlprefix{URL }\fi
\providecommand{\bibinfo}[2]{#2}
\providecommand{\eprint}[2][]{\url{#2}}

\bibitem[{\citenamefont{Carr et~al.}(2009)\citenamefont{Carr, DeMille, Krems,
  and Ye}}]{Carr:2009}
\bibinfo{author}{\bibfnamefont{L.~D.} \bibnamefont{Carr}},
  \bibinfo{author}{\bibfnamefont{D.}~\bibnamefont{DeMille}},
  \bibinfo{author}{\bibfnamefont{R.~V.} \bibnamefont{Krems}}, \bibnamefont{and}
  \bibinfo{author}{\bibfnamefont{J.}~\bibnamefont{Ye}}, \bibinfo{journal}{New
  Journal of Physics} \textbf{\bibinfo{volume}{11}}, \bibinfo{pages}{055049}
  (\bibinfo{year}{2009}),
  \urlprefix\url{https://doi.org/10.1088%2F1367-2630%2F11%2F5%2F055049}.

\bibitem[{\citenamefont{Zelevinsky et~al.}(2008)\citenamefont{Zelevinsky,
  Kotochigova, and Ye}}]{Zelevinsky:2008}
\bibinfo{author}{\bibfnamefont{T.}~\bibnamefont{Zelevinsky}},
  \bibinfo{author}{\bibfnamefont{S.}~\bibnamefont{Kotochigova}},
  \bibnamefont{and} \bibinfo{author}{\bibfnamefont{J.}~\bibnamefont{Ye}},
  \bibinfo{journal}{Physical Review Letters} \textbf{\bibinfo{volume}{100}},
  \bibinfo{pages}{043201} (\bibinfo{year}{2008}),
  \urlprefix\url{https://doi.org/10.1103%2Fphysrevlett.100.043201}.

\bibitem[{\citenamefont{Salumbides et~al.}(2011)\citenamefont{Salumbides,
  Dickenson, Ivanov, and Ubachs}}]{Salumbides:2011}
\bibinfo{author}{\bibfnamefont{E.~J.} \bibnamefont{Salumbides}},
  \bibinfo{author}{\bibfnamefont{G.~D.} \bibnamefont{Dickenson}},
  \bibinfo{author}{\bibfnamefont{T.~I.} \bibnamefont{Ivanov}},
  \bibnamefont{and} \bibinfo{author}{\bibfnamefont{W.}~\bibnamefont{Ubachs}},
  \bibinfo{journal}{Physical Review Letters} \textbf{\bibinfo{volume}{107}},
  \bibinfo{pages}{043005} (\bibinfo{year}{2011}),
  \urlprefix\url{https://doi.org/10.1103%2Fphysrevlett.107.043005}.

\bibitem[{\citenamefont{Salumbides et~al.}(2013)\citenamefont{Salumbides,
  Koelemeij, Komasa, Pachucki, Eikema, and Ubachs}}]{Salumbides:2013}
\bibinfo{author}{\bibfnamefont{E.~J.} \bibnamefont{Salumbides}},
  \bibinfo{author}{\bibfnamefont{J.~C.~J.} \bibnamefont{Koelemeij}},
  \bibinfo{author}{\bibfnamefont{J.}~\bibnamefont{Komasa}},
  \bibinfo{author}{\bibfnamefont{K.}~\bibnamefont{Pachucki}},
  \bibinfo{author}{\bibfnamefont{K.~S.~E.} \bibnamefont{Eikema}},
  \bibnamefont{and} \bibinfo{author}{\bibfnamefont{W.}~\bibnamefont{Ubachs}},
  \bibinfo{journal}{Physical Review D} \textbf{\bibinfo{volume}{87}},
  \bibinfo{pages}{112008} (\bibinfo{year}{2013}),
  \urlprefix\url{https://doi.org/10.1103%2Fphysrevd.87.112008}.

\bibitem[{\citenamefont{Tarbutt et~al.}(2013)\citenamefont{Tarbutt, Sauer,
  Hudson, and Hinds}}]{Tarbutt:2013}
\bibinfo{author}{\bibfnamefont{M.~R.} \bibnamefont{Tarbutt}},
  \bibinfo{author}{\bibfnamefont{B.~E.} \bibnamefont{Sauer}},
  \bibinfo{author}{\bibfnamefont{J.~J.} \bibnamefont{Hudson}},
  \bibnamefont{and} \bibinfo{author}{\bibfnamefont{E.~A.} \bibnamefont{Hinds}},
  \bibinfo{journal}{New Journal of Physics} \textbf{\bibinfo{volume}{15}},
  \bibinfo{pages}{053034} (\bibinfo{year}{2013}),
  \urlprefix\url{https://doi.org/10.1088%2F1367-2630%2F15%2F5%2F053034}.

\bibitem[{\citenamefont{Schiller et~al.}(2014)\citenamefont{Schiller, Bakalov,
  and Korobov}}]{Schiller:2014}
\bibinfo{author}{\bibfnamefont{S.}~\bibnamefont{Schiller}},
  \bibinfo{author}{\bibfnamefont{D.}~\bibnamefont{Bakalov}}, \bibnamefont{and}
  \bibinfo{author}{\bibfnamefont{V.}~\bibnamefont{Korobov}},
  \bibinfo{journal}{Physical Review Letters} \textbf{\bibinfo{volume}{113}},
  \bibinfo{pages}{023004} (\bibinfo{year}{2014}),
  \urlprefix\url{https://doi.org/10.1103%2Fphysrevlett.113.023004}.

\bibitem[{\citenamefont{Borkowski}(2018)}]{Borkowski:2018}
\bibinfo{author}{\bibfnamefont{M.}~\bibnamefont{Borkowski}},
  \bibinfo{journal}{Physical Review Letters} \textbf{\bibinfo{volume}{120}},
  \bibinfo{pages}{083202} (\bibinfo{year}{2018}),
  \urlprefix\url{https://doi.org/10.1103%2Fphysrevlett.120.083202}.

\bibitem[{\citenamefont{Borkowski et~al.}(2019)\citenamefont{Borkowski,
  Buchachenko, Ciuryło, Julienne, Yamada, Kikuchi, Takasu, and
  Takahashi}}]{Borkowski:2019}
\bibinfo{author}{\bibfnamefont{M.}~\bibnamefont{Borkowski}},
  \bibinfo{author}{\bibfnamefont{A.~A.} \bibnamefont{Buchachenko}},
  \bibinfo{author}{\bibfnamefont{R.}~\bibnamefont{Ciuryło}},
  \bibinfo{author}{\bibfnamefont{P.~S.} \bibnamefont{Julienne}},
  \bibinfo{author}{\bibfnamefont{H.}~\bibnamefont{Yamada}},
  \bibinfo{author}{\bibfnamefont{Y.}~\bibnamefont{Kikuchi}},
  \bibinfo{author}{\bibfnamefont{Y.}~\bibnamefont{Takasu}}, \bibnamefont{and}
  \bibinfo{author}{\bibfnamefont{Y.}~\bibnamefont{Takahashi}},
  \bibinfo{journal}{Scientific Reports} \textbf{\bibinfo{volume}{9}}
  (\bibinfo{year}{2019}), ISSN \bibinfo{issn}{2045-2322}.

\bibitem[{\citenamefont{Krems}(2008)}]{Krems:2008}
\bibinfo{author}{\bibfnamefont{R.~V.} \bibnamefont{Krems}},
  \bibinfo{journal}{Physical Chemistry Chemical Physics}
  \textbf{\bibinfo{volume}{10}}, \bibinfo{pages}{4079} (\bibinfo{year}{2008}).

\bibitem[{\citenamefont{Bell and Softley}(2009)}]{Bell:2009}
\bibinfo{author}{\bibfnamefont{M.~T.} \bibnamefont{Bell}} \bibnamefont{and}
  \bibinfo{author}{\bibfnamefont{T.~P.} \bibnamefont{Softley}},
  \bibinfo{journal}{Molecular Physics} \textbf{\bibinfo{volume}{107}},
  \bibinfo{pages}{99} (\bibinfo{year}{2009}).

\bibitem[{\citenamefont{Ospelkaus
  et~al.}(2010{\natexlab{a}})\citenamefont{Ospelkaus, Ni, Wang, de~Miranda,
  Neyenhuis, Qu{\'e}m{\'e}ner, Julienne, Bohn, Jin, and Ye}}]{Ospelkaus:2010}
\bibinfo{author}{\bibfnamefont{S.}~\bibnamefont{Ospelkaus}},
  \bibinfo{author}{\bibfnamefont{K.-K.} \bibnamefont{Ni}},
  \bibinfo{author}{\bibfnamefont{D.}~\bibnamefont{Wang}},
  \bibinfo{author}{\bibfnamefont{M.~H.~G.} \bibnamefont{de~Miranda}},
  \bibinfo{author}{\bibfnamefont{B.}~\bibnamefont{Neyenhuis}},
  \bibinfo{author}{\bibfnamefont{G.}~\bibnamefont{Qu{\'e}m{\'e}ner}},
  \bibinfo{author}{\bibfnamefont{P.~S.} \bibnamefont{Julienne}},
  \bibinfo{author}{\bibfnamefont{J.~L.} \bibnamefont{Bohn}},
  \bibinfo{author}{\bibfnamefont{D.~S.} \bibnamefont{Jin}}, \bibnamefont{and}
  \bibinfo{author}{\bibfnamefont{J.}~\bibnamefont{Ye}},
  \bibinfo{journal}{Science} \textbf{\bibinfo{volume}{327}},
  \bibinfo{pages}{853} (\bibinfo{year}{2010}{\natexlab{a}}), ISSN
  \bibinfo{issn}{0036-8075},
  \urlprefix\url{https://science.sciencemag.org/content/327/5967/853}.

\bibitem[{\citenamefont{Dulieu et~al.}(2011)\citenamefont{Dulieu, Krems,
  Weidem\"{u}ller, and Willitsch}}]{Dulieu:2011}
\bibinfo{author}{\bibfnamefont{O.}~\bibnamefont{Dulieu}},
  \bibinfo{author}{\bibfnamefont{R.}~\bibnamefont{Krems}},
  \bibinfo{author}{\bibfnamefont{M.}~\bibnamefont{Weidem\"{u}ller}},
  \bibnamefont{and}
  \bibinfo{author}{\bibfnamefont{S.}~\bibnamefont{Willitsch}},
  \bibinfo{journal}{Physical Chemistry Chemical Physics}
  \textbf{\bibinfo{volume}{13}}, \bibinfo{pages}{18703} (\bibinfo{year}{2011}).

\bibitem[{\citenamefont{Balakrishnan}(2016)}]{Balakrishnan:2016}
\bibinfo{author}{\bibfnamefont{N.}~\bibnamefont{Balakrishnan}},
  \bibinfo{journal}{Journal of Chemical Physics}
  \textbf{\bibinfo{volume}{145}}, \bibinfo{pages}{150901}
  (\bibinfo{year}{2016}).

\bibitem[{\citenamefont{Santos et~al.}(2000)\citenamefont{Santos, Shlyapnikov,
  Zoller, and Lewenstein}}]{Santos:2000}
\bibinfo{author}{\bibfnamefont{L.}~\bibnamefont{Santos}},
  \bibinfo{author}{\bibfnamefont{G.~V.} \bibnamefont{Shlyapnikov}},
  \bibinfo{author}{\bibfnamefont{P.}~\bibnamefont{Zoller}}, \bibnamefont{and}
  \bibinfo{author}{\bibfnamefont{M.}~\bibnamefont{Lewenstein}},
  \bibinfo{journal}{Physical Review Letters} \textbf{\bibinfo{volume}{85}},
  \bibinfo{pages}{1791} (\bibinfo{year}{2000}).

\bibitem[{\citenamefont{Micheli et~al.}(2007)\citenamefont{Micheli, Pupillo,
  Büchler, and Zoller}}]{Micheli:2007}
\bibinfo{author}{\bibfnamefont{A.}~\bibnamefont{Micheli}},
  \bibinfo{author}{\bibfnamefont{G.}~\bibnamefont{Pupillo}},
  \bibinfo{author}{\bibfnamefont{H.~P.} \bibnamefont{Büchler}},
  \bibnamefont{and} \bibinfo{author}{\bibfnamefont{P.}~\bibnamefont{Zoller}},
  \bibinfo{journal}{Physical Review A} \textbf{\bibinfo{volume}{76}},
  \bibinfo{pages}{043604} (\bibinfo{year}{2007}),
  \urlprefix\url{https://doi.org/10.1103%2Fphysreva.76.043604}.

\bibitem[{\citenamefont{Pollet et~al.}(2010)\citenamefont{Pollet, Picon,
  Büchler, and Troyer}}]{Pollet:2010}
\bibinfo{author}{\bibfnamefont{L.}~\bibnamefont{Pollet}},
  \bibinfo{author}{\bibfnamefont{J.~D.} \bibnamefont{Picon}},
  \bibinfo{author}{\bibfnamefont{H.~P.} \bibnamefont{Büchler}},
  \bibnamefont{and} \bibinfo{author}{\bibfnamefont{M.}~\bibnamefont{Troyer}},
  \bibinfo{journal}{Physical Review Letters} \textbf{\bibinfo{volume}{104}},
  \bibinfo{pages}{125302} (\bibinfo{year}{2010}),
  \urlprefix\url{https://doi.org/10.1103%2Fphysrevlett.104.125302}.

\bibitem[{\citenamefont{Capogrosso-Sansone
  et~al.}(2010)\citenamefont{Capogrosso-Sansone, Trefzger, Lewenstein, Zoller,
  and Pupillo}}]{Capogrosso_Sansone:2010}
\bibinfo{author}{\bibfnamefont{B.}~\bibnamefont{Capogrosso-Sansone}},
  \bibinfo{author}{\bibfnamefont{C.}~\bibnamefont{Trefzger}},
  \bibinfo{author}{\bibfnamefont{M.}~\bibnamefont{Lewenstein}},
  \bibinfo{author}{\bibfnamefont{P.}~\bibnamefont{Zoller}}, \bibnamefont{and}
  \bibinfo{author}{\bibfnamefont{G.}~\bibnamefont{Pupillo}},
  \bibinfo{journal}{Physical Review Letters} \textbf{\bibinfo{volume}{104}},
  \bibinfo{pages}{125301} (\bibinfo{year}{2010}),
  \urlprefix\url{https://doi.org/10.1103%2Fphysrevlett.104.125301}.

\bibitem[{\citenamefont{Baranov et~al.}(2012)\citenamefont{Baranov, Dalmonte,
  Pupillo, and Zoller}}]{Baranov:2012}
\bibinfo{author}{\bibfnamefont{M.~A.} \bibnamefont{Baranov}},
  \bibinfo{author}{\bibfnamefont{M.}~\bibnamefont{Dalmonte}},
  \bibinfo{author}{\bibfnamefont{G.}~\bibnamefont{Pupillo}}, \bibnamefont{and}
  \bibinfo{author}{\bibfnamefont{P.}~\bibnamefont{Zoller}},
  \bibinfo{journal}{Chemical Reviews} \textbf{\bibinfo{volume}{112}},
  \bibinfo{pages}{5012} (\bibinfo{year}{2012}).

\bibitem[{\citenamefont{Lechner and Zoller}(2013)}]{Lechner:2013}
\bibinfo{author}{\bibfnamefont{W.}~\bibnamefont{Lechner}} \bibnamefont{and}
  \bibinfo{author}{\bibfnamefont{P.}~\bibnamefont{Zoller}},
  \bibinfo{journal}{Physical Review Letters} \textbf{\bibinfo{volume}{111}},
  \bibinfo{pages}{185306} (\bibinfo{year}{2013}),
  \urlprefix\url{https://doi.org/10.1103%2Fphysrevlett.111.185306}.

\bibitem[{\citenamefont{Barnett et~al.}(2006)\citenamefont{Barnett, Petrov,
  Lukin, and Demler}}]{Barnett:2006}
\bibinfo{author}{\bibfnamefont{R.}~\bibnamefont{Barnett}},
  \bibinfo{author}{\bibfnamefont{D.}~\bibnamefont{Petrov}},
  \bibinfo{author}{\bibfnamefont{M.}~\bibnamefont{Lukin}}, \bibnamefont{and}
  \bibinfo{author}{\bibfnamefont{E.}~\bibnamefont{Demler}},
  \bibinfo{journal}{Physical Review Letters} \textbf{\bibinfo{volume}{96}},
  \bibinfo{pages}{190401} (\bibinfo{year}{2006}).

\bibitem[{\citenamefont{Micheli et~al.}(2006)\citenamefont{Micheli, Brennen,
  and Zoller}}]{Micheli:2006}
\bibinfo{author}{\bibfnamefont{A.}~\bibnamefont{Micheli}},
  \bibinfo{author}{\bibfnamefont{G.~K.} \bibnamefont{Brennen}},
  \bibnamefont{and} \bibinfo{author}{\bibfnamefont{P.}~\bibnamefont{Zoller}},
  \bibinfo{journal}{Nature Physics} \textbf{\bibinfo{volume}{2}},
  \bibinfo{pages}{341} (\bibinfo{year}{2006}).

\bibitem[{\citenamefont{B\"{u}chler et~al.}(2007)\citenamefont{B\"{u}chler,
  Demler, Lukin, Micheli, Prokof'ev, Pupillo, and Zoller}}]{Buchler:2007}
\bibinfo{author}{\bibfnamefont{H.~P.} \bibnamefont{B\"{u}chler}},
  \bibinfo{author}{\bibfnamefont{E.}~\bibnamefont{Demler}},
  \bibinfo{author}{\bibfnamefont{M.}~\bibnamefont{Lukin}},
  \bibinfo{author}{\bibfnamefont{A.}~\bibnamefont{Micheli}},
  \bibinfo{author}{\bibfnamefont{N.}~\bibnamefont{Prokof'ev}},
  \bibinfo{author}{\bibfnamefont{G.}~\bibnamefont{Pupillo}}, \bibnamefont{and}
  \bibinfo{author}{\bibfnamefont{P.}~\bibnamefont{Zoller}},
  \bibinfo{journal}{Physical Review Letters} \textbf{\bibinfo{volume}{98}},
  \bibinfo{pages}{060404} (\bibinfo{year}{2007}).

\bibitem[{\citenamefont{Maci{\`{a}} et~al.}(2012)\citenamefont{Maci{\`{a}},
  Hufnagl, Mazzanti, Boronat, and Zillich}}]{Macia:2012}
\bibinfo{author}{\bibfnamefont{A.}~\bibnamefont{Maci{\`{a}}}},
  \bibinfo{author}{\bibfnamefont{D.}~\bibnamefont{Hufnagl}},
  \bibinfo{author}{\bibfnamefont{F.}~\bibnamefont{Mazzanti}},
  \bibinfo{author}{\bibfnamefont{J.}~\bibnamefont{Boronat}}, \bibnamefont{and}
  \bibinfo{author}{\bibfnamefont{R.~E.} \bibnamefont{Zillich}},
  \bibinfo{journal}{Physical Review Letters} \textbf{\bibinfo{volume}{109}},
  \bibinfo{pages}{235307} (\bibinfo{year}{2012}).

\bibitem[{\citenamefont{Manmana et~al.}(2013)\citenamefont{Manmana,
  Stoudenmire, Hazzard, Rey, and Gorshkov}}]{Manmana:2013}
\bibinfo{author}{\bibfnamefont{S.~R.} \bibnamefont{Manmana}},
  \bibinfo{author}{\bibfnamefont{E.~M.} \bibnamefont{Stoudenmire}},
  \bibinfo{author}{\bibfnamefont{K.~R.~A.} \bibnamefont{Hazzard}},
  \bibinfo{author}{\bibfnamefont{A.~M.} \bibnamefont{Rey}}, \bibnamefont{and}
  \bibinfo{author}{\bibfnamefont{A.~V.} \bibnamefont{Gorshkov}},
  \bibinfo{journal}{Physical Review B} \textbf{\bibinfo{volume}{87}},
  \bibinfo{pages}{081106} (\bibinfo{year}{2013}),
  \urlprefix\url{https://doi.org/10.1103%2Fphysrevb.87.081106}.

\bibitem[{\citenamefont{Gorshkov et~al.}(2013)\citenamefont{Gorshkov, Hazzard,
  and Rey}}]{Gorshkov:2013}
\bibinfo{author}{\bibfnamefont{A.~V.} \bibnamefont{Gorshkov}},
  \bibinfo{author}{\bibfnamefont{K.~R.~A.} \bibnamefont{Hazzard}},
  \bibnamefont{and} \bibinfo{author}{\bibfnamefont{A.~M.} \bibnamefont{Rey}},
  \bibinfo{journal}{Molecular Physics} \textbf{\bibinfo{volume}{111}},
  \bibinfo{pages}{1908} (\bibinfo{year}{2013}), ISSN \bibinfo{issn}{0026-8976}.

\bibitem[{\citenamefont{DeMille}(2002)}]{Demille:2002}
\bibinfo{author}{\bibfnamefont{D.}~\bibnamefont{DeMille}},
  \bibinfo{journal}{Physical Review Letters} \textbf{\bibinfo{volume}{88}},
  \bibinfo{pages}{067901} (\bibinfo{year}{2002}),
  \urlprefix\url{https://link.aps.org/doi/10.1103/PhysRevLett.88.067901}.

\bibitem[{\citenamefont{Yelin et~al.}(2006)\citenamefont{Yelin, Kirby, and
  C{\^{o}}t{\'{e}}}}]{Yelin:2006}
\bibinfo{author}{\bibfnamefont{S.~F.} \bibnamefont{Yelin}},
  \bibinfo{author}{\bibfnamefont{K.}~\bibnamefont{Kirby}}, \bibnamefont{and}
  \bibinfo{author}{\bibfnamefont{R.}~\bibnamefont{C{\^{o}}t{\'{e}}}},
  \bibinfo{journal}{Physical Review A} \textbf{\bibinfo{volume}{74}},
  \bibinfo{pages}{050301} (\bibinfo{year}{2006}),
  \urlprefix\url{https://doi.org/10.1103%2Fphysreva.74.050301}.

\bibitem[{\citenamefont{Zhu et~al.}(2013)\citenamefont{Zhu, Kais, Wei,
  Herschbach, and Friedrich}}]{Zhu:2013}
\bibinfo{author}{\bibfnamefont{J.}~\bibnamefont{Zhu}},
  \bibinfo{author}{\bibfnamefont{S.}~\bibnamefont{Kais}},
  \bibinfo{author}{\bibfnamefont{Q.}~\bibnamefont{Wei}},
  \bibinfo{author}{\bibfnamefont{D.}~\bibnamefont{Herschbach}},
  \bibnamefont{and}
  \bibinfo{author}{\bibfnamefont{B.}~\bibnamefont{Friedrich}},
  \bibinfo{journal}{The Journal of Chemical Physics}
  \textbf{\bibinfo{volume}{138}}, \bibinfo{pages}{024104}
  (\bibinfo{year}{2013}), ISSN \bibinfo{issn}{0021-9606}.

\bibitem[{\citenamefont{Herrera et~al.}(2014)\citenamefont{Herrera, Cao, Kais,
  and Whaley}}]{Herrera:2014}
\bibinfo{author}{\bibfnamefont{F.}~\bibnamefont{Herrera}},
  \bibinfo{author}{\bibfnamefont{Y.}~\bibnamefont{Cao}},
  \bibinfo{author}{\bibfnamefont{S.}~\bibnamefont{Kais}}, \bibnamefont{and}
  \bibinfo{author}{\bibfnamefont{K.~B.} \bibnamefont{Whaley}},
  \bibinfo{journal}{New Journal of Physics} \textbf{\bibinfo{volume}{16}},
  \bibinfo{pages}{075001} (\bibinfo{year}{2014}),
  \urlprefix\url{https://doi.org/10.1088%2F1367-2630%2F16%2F7%2F075001}.

\bibitem[{\citenamefont{Ni et~al.}(2018)\citenamefont{Ni, Rosenband, and
  Grimes}}]{Ni:2018}
\bibinfo{author}{\bibfnamefont{K.-K.} \bibnamefont{Ni}},
  \bibinfo{author}{\bibfnamefont{T.}~\bibnamefont{Rosenband}},
  \bibnamefont{and} \bibinfo{author}{\bibfnamefont{D.~D.}
  \bibnamefont{Grimes}}, \bibinfo{journal}{Chemical Science}
  \textbf{\bibinfo{volume}{9}}, \bibinfo{pages}{6830} (\bibinfo{year}{2018}),
  \urlprefix\url{http://dx.doi.org/10.1039/C8SC02355G}.

\bibitem[{\citenamefont{Sawant et~al.}(2020)\citenamefont{Sawant, Blackmore,
  Gregory, Mur-Petit, Jaksch, Aldegunde, Hutson, Tarbutt, and
  Cornish}}]{Sawant:2020}
\bibinfo{author}{\bibfnamefont{R.}~\bibnamefont{Sawant}},
  \bibinfo{author}{\bibfnamefont{J.~A.} \bibnamefont{Blackmore}},
  \bibinfo{author}{\bibfnamefont{P.~D.} \bibnamefont{Gregory}},
  \bibinfo{author}{\bibfnamefont{J.}~\bibnamefont{Mur-Petit}},
  \bibinfo{author}{\bibfnamefont{D.}~\bibnamefont{Jaksch}},
  \bibinfo{author}{\bibfnamefont{J.}~\bibnamefont{Aldegunde}},
  \bibinfo{author}{\bibfnamefont{J.~M.} \bibnamefont{Hutson}},
  \bibinfo{author}{\bibfnamefont{M.~R.} \bibnamefont{Tarbutt}},
  \bibnamefont{and} \bibinfo{author}{\bibfnamefont{S.~L.}
  \bibnamefont{Cornish}}, \bibinfo{journal}{New Journal of Physics}
  \textbf{\bibinfo{volume}{22}}, \bibinfo{pages}{013027}
  (\bibinfo{year}{2020}),
  \urlprefix\url{https://doi.org/10.1088%2F1367-2630%2Fab60f4}.

\bibitem[{\citenamefont{Hughes et~al.}(2020)\citenamefont{Hughes, Frye, Sawant,
  Bhole, Jones, Cornish, Tarbutt, Hutson, Jaksch, and Mur-Petit}}]{Hughes:2019}
\bibinfo{author}{\bibfnamefont{M.}~\bibnamefont{Hughes}},
  \bibinfo{author}{\bibfnamefont{M.~D.} \bibnamefont{Frye}},
  \bibinfo{author}{\bibfnamefont{R.}~\bibnamefont{Sawant}},
  \bibinfo{author}{\bibfnamefont{G.}~\bibnamefont{Bhole}},
  \bibinfo{author}{\bibfnamefont{J.~A.} \bibnamefont{Jones}},
  \bibinfo{author}{\bibfnamefont{S.~L.} \bibnamefont{Cornish}},
  \bibinfo{author}{\bibfnamefont{M.~R.} \bibnamefont{Tarbutt}},
  \bibinfo{author}{\bibfnamefont{J.~M.} \bibnamefont{Hutson}},
  \bibinfo{author}{\bibfnamefont{D.}~\bibnamefont{Jaksch}}, \bibnamefont{and}
  \bibinfo{author}{\bibfnamefont{J.}~\bibnamefont{Mur-Petit}},
  \bibinfo{journal}{Phys. Rev. A} \textbf{\bibinfo{volume}{101}},
  \bibinfo{pages}{062308} (\bibinfo{year}{2020}),
  \urlprefix\url{https://link.aps.org/doi/10.1103/PhysRevA.101.062308}.

\bibitem[{\citenamefont{Ni et~al.}(2008)\citenamefont{Ni, Ospelkaus,
  de~Miranda, Pe'er, Neyenhuis, Zirbel, Kotochigova, Julienne, Jin, and
  Ye}}]{Ni:2008}
\bibinfo{author}{\bibfnamefont{K.-K.} \bibnamefont{Ni}},
  \bibinfo{author}{\bibfnamefont{S.}~\bibnamefont{Ospelkaus}},
  \bibinfo{author}{\bibfnamefont{M.~H.~G.} \bibnamefont{de~Miranda}},
  \bibinfo{author}{\bibfnamefont{A.}~\bibnamefont{Pe'er}},
  \bibinfo{author}{\bibfnamefont{B.}~\bibnamefont{Neyenhuis}},
  \bibinfo{author}{\bibfnamefont{J.~J.} \bibnamefont{Zirbel}},
  \bibinfo{author}{\bibfnamefont{S.}~\bibnamefont{Kotochigova}},
  \bibinfo{author}{\bibfnamefont{P.~S.} \bibnamefont{Julienne}},
  \bibinfo{author}{\bibfnamefont{D.~S.} \bibnamefont{Jin}}, \bibnamefont{and}
  \bibinfo{author}{\bibfnamefont{J.}~\bibnamefont{Ye}},
  \bibinfo{journal}{Science} \textbf{\bibinfo{volume}{322}},
  \bibinfo{pages}{231} (\bibinfo{year}{2008}).

\bibitem[{\citenamefont{Danzl et~al.}(2008)\citenamefont{Danzl, Haller,
  Gustavsson, Mark, Hart, Bouloufa, Dulieu, Ritsch, and
  N\"{a}gerl}}]{Danzl:2008}
\bibinfo{author}{\bibfnamefont{J.~G.} \bibnamefont{Danzl}},
  \bibinfo{author}{\bibfnamefont{E.}~\bibnamefont{Haller}},
  \bibinfo{author}{\bibfnamefont{M.}~\bibnamefont{Gustavsson}},
  \bibinfo{author}{\bibfnamefont{M.~J.} \bibnamefont{Mark}},
  \bibinfo{author}{\bibfnamefont{R.}~\bibnamefont{Hart}},
  \bibinfo{author}{\bibfnamefont{N.}~\bibnamefont{Bouloufa}},
  \bibinfo{author}{\bibfnamefont{O.}~\bibnamefont{Dulieu}},
  \bibinfo{author}{\bibfnamefont{H.}~\bibnamefont{Ritsch}}, \bibnamefont{and}
  \bibinfo{author}{\bibfnamefont{H.-C.} \bibnamefont{N\"{a}gerl}},
  \bibinfo{journal}{Science} \textbf{\bibinfo{volume}{321}},
  \bibinfo{pages}{1062} (\bibinfo{year}{2008}).

\bibitem[{\citenamefont{Lang et~al.}(2008)\citenamefont{Lang, Winkler, Strauss,
  Grimm, and Hecker~Denschlag}}]{Lang:2008}
\bibinfo{author}{\bibfnamefont{F.}~\bibnamefont{Lang}},
  \bibinfo{author}{\bibfnamefont{K.}~\bibnamefont{Winkler}},
  \bibinfo{author}{\bibfnamefont{C.}~\bibnamefont{Strauss}},
  \bibinfo{author}{\bibfnamefont{R.}~\bibnamefont{Grimm}}, \bibnamefont{and}
  \bibinfo{author}{\bibfnamefont{J.}~\bibnamefont{Hecker~Denschlag}},
  \bibinfo{journal}{Physical Review Letters} \textbf{\bibinfo{volume}{101}},
  \bibinfo{pages}{133005} (\bibinfo{year}{2008}).

\bibitem[{\citenamefont{Takekoshi et~al.}(2014)\citenamefont{Takekoshi,
  Reichs\"{o}llner, Schindewolf, Hutson, Le~Sueur, Dulieu, Ferlaino, Grimm, and
  N\"{a}gerl}}]{Takekoshi:2014}
\bibinfo{author}{\bibfnamefont{T.}~\bibnamefont{Takekoshi}},
  \bibinfo{author}{\bibfnamefont{L.}~\bibnamefont{Reichs\"{o}llner}},
  \bibinfo{author}{\bibfnamefont{A.}~\bibnamefont{Schindewolf}},
  \bibinfo{author}{\bibfnamefont{J.~M.} \bibnamefont{Hutson}},
  \bibinfo{author}{\bibfnamefont{C.~R.} \bibnamefont{Le~Sueur}},
  \bibinfo{author}{\bibfnamefont{O.}~\bibnamefont{Dulieu}},
  \bibinfo{author}{\bibfnamefont{F.}~\bibnamefont{Ferlaino}},
  \bibinfo{author}{\bibfnamefont{R.}~\bibnamefont{Grimm}}, \bibnamefont{and}
  \bibinfo{author}{\bibfnamefont{H.-C.} \bibnamefont{N\"{a}gerl}},
  \bibinfo{journal}{Physical Review Letters} \textbf{\bibinfo{volume}{113}},
  \bibinfo{pages}{205301} (\bibinfo{year}{2014}).

\bibitem[{\citenamefont{Molony et~al.}(2014)\citenamefont{Molony, Gregory, Ji,
  Lu, K\"{o}ppinger, Le~Sueur, Blackley, Hutson, and Cornish}}]{Molony:2014}
\bibinfo{author}{\bibfnamefont{P.~K.} \bibnamefont{Molony}},
  \bibinfo{author}{\bibfnamefont{P.~D.} \bibnamefont{Gregory}},
  \bibinfo{author}{\bibfnamefont{Z.}~\bibnamefont{Ji}},
  \bibinfo{author}{\bibfnamefont{B.}~\bibnamefont{Lu}},
  \bibinfo{author}{\bibfnamefont{M.~P.} \bibnamefont{K\"{o}ppinger}},
  \bibinfo{author}{\bibfnamefont{C.~R.} \bibnamefont{Le~Sueur}},
  \bibinfo{author}{\bibfnamefont{C.~L.} \bibnamefont{Blackley}},
  \bibinfo{author}{\bibfnamefont{J.~M.} \bibnamefont{Hutson}},
  \bibnamefont{and} \bibinfo{author}{\bibfnamefont{S.~L.}
  \bibnamefont{Cornish}}, \bibinfo{journal}{Physical Review Letters}
  \textbf{\bibinfo{volume}{113}}, \bibinfo{pages}{255301}
  (\bibinfo{year}{2014}).

\bibitem[{\citenamefont{Park et~al.}(2015)\citenamefont{Park, Will, and
  Zwierlein}}]{Park:2015}
\bibinfo{author}{\bibfnamefont{J.~W.} \bibnamefont{Park}},
  \bibinfo{author}{\bibfnamefont{S.~A.} \bibnamefont{Will}}, \bibnamefont{and}
  \bibinfo{author}{\bibfnamefont{M.~W.} \bibnamefont{Zwierlein}},
  \bibinfo{journal}{Physical Review Letters} \textbf{\bibinfo{volume}{114}},
  \bibinfo{pages}{205302} (\bibinfo{year}{2015}).

\bibitem[{\citenamefont{Guo et~al.}(2016)\citenamefont{Guo, Zhu, Lu, Ye, Wang,
  Vexiau, {Bouloufa-Maafa}, Qu\'{e}m\'{e}ner, Dulieu, and Wang}}]{Guo:2016}
\bibinfo{author}{\bibfnamefont{M.}~\bibnamefont{Guo}},
  \bibinfo{author}{\bibfnamefont{B.}~\bibnamefont{Zhu}},
  \bibinfo{author}{\bibfnamefont{B.}~\bibnamefont{Lu}},
  \bibinfo{author}{\bibfnamefont{X.}~\bibnamefont{Ye}},
  \bibinfo{author}{\bibfnamefont{F.}~\bibnamefont{Wang}},
  \bibinfo{author}{\bibfnamefont{R.}~\bibnamefont{Vexiau}},
  \bibinfo{author}{\bibfnamefont{N.}~\bibnamefont{{Bouloufa-Maafa}}},
  \bibinfo{author}{\bibfnamefont{G.}~\bibnamefont{Qu\'{e}m\'{e}ner}},
  \bibinfo{author}{\bibfnamefont{O.}~\bibnamefont{Dulieu}}, \bibnamefont{and}
  \bibinfo{author}{\bibfnamefont{D.}~\bibnamefont{Wang}},
  \bibinfo{journal}{Physical Review Letters} \textbf{\bibinfo{volume}{116}},
  \bibinfo{pages}{205303} (\bibinfo{year}{2016}).

\bibitem[{\citenamefont{Rvachov et~al.}(2017)\citenamefont{Rvachov, Son,
  Sommer, Ebadi, Park, Zwierlein, Ketterle, and Jamison}}]{Rvachov:2017}
\bibinfo{author}{\bibfnamefont{T.~M.} \bibnamefont{Rvachov}},
  \bibinfo{author}{\bibfnamefont{H.}~\bibnamefont{Son}},
  \bibinfo{author}{\bibfnamefont{A.~T.} \bibnamefont{Sommer}},
  \bibinfo{author}{\bibfnamefont{S.}~\bibnamefont{Ebadi}},
  \bibinfo{author}{\bibfnamefont{J.~J.} \bibnamefont{Park}},
  \bibinfo{author}{\bibfnamefont{M.~W.} \bibnamefont{Zwierlein}},
  \bibinfo{author}{\bibfnamefont{W.}~\bibnamefont{Ketterle}}, \bibnamefont{and}
  \bibinfo{author}{\bibfnamefont{A.~O.} \bibnamefont{Jamison}},
  \bibinfo{journal}{Physical Review Letters} \textbf{\bibinfo{volume}{119}},
  \bibinfo{pages}{143001} (\bibinfo{year}{2017}).

\bibitem[{\citenamefont{See{\ss}elberg
  et~al.}(2018)\citenamefont{See{\ss}elberg, Buchheim, Lu, Schneider, Luo,
  Tiemann, Bloch, and Gohle}}]{Seesselberg:2018b}
\bibinfo{author}{\bibfnamefont{F.}~\bibnamefont{See{\ss}elberg}},
  \bibinfo{author}{\bibfnamefont{N.}~\bibnamefont{Buchheim}},
  \bibinfo{author}{\bibfnamefont{Z.-K.} \bibnamefont{Lu}},
  \bibinfo{author}{\bibfnamefont{T.}~\bibnamefont{Schneider}},
  \bibinfo{author}{\bibfnamefont{X.-Y.} \bibnamefont{Luo}},
  \bibinfo{author}{\bibfnamefont{E.}~\bibnamefont{Tiemann}},
  \bibinfo{author}{\bibfnamefont{I.}~\bibnamefont{Bloch}}, \bibnamefont{and}
  \bibinfo{author}{\bibfnamefont{C.}~\bibnamefont{Gohle}},
  \bibinfo{journal}{Physical Review A} \textbf{\bibinfo{volume}{97}},
  \bibinfo{pages}{013405} (\bibinfo{year}{2018}),
  \urlprefix\url{https://doi.org/10.1103%2Fphysreva.97.013405}.

\bibitem[{\citenamefont{Yang et~al.}(2019)\citenamefont{Yang, Zhang, Liu, Liu,
  Nan, Zhao, and Pan}}]{Yang:2019}
\bibinfo{author}{\bibfnamefont{H.}~\bibnamefont{Yang}},
  \bibinfo{author}{\bibfnamefont{D.-C.} \bibnamefont{Zhang}},
  \bibinfo{author}{\bibfnamefont{L.}~\bibnamefont{Liu}},
  \bibinfo{author}{\bibfnamefont{Y.-X.} \bibnamefont{Liu}},
  \bibinfo{author}{\bibfnamefont{J.}~\bibnamefont{Nan}},
  \bibinfo{author}{\bibfnamefont{B.}~\bibnamefont{Zhao}}, \bibnamefont{and}
  \bibinfo{author}{\bibfnamefont{J.-W.} \bibnamefont{Pan}},
  \bibinfo{journal}{Science} \textbf{\bibinfo{volume}{363}},
  \bibinfo{pages}{261} (\bibinfo{year}{2019}), ISSN \bibinfo{issn}{0036-8075},
  \urlprefix\url{https://science.sciencemag.org/content/363/6424/261}.

\bibitem[{\citenamefont{Voges et~al.}(2020)\citenamefont{Voges, Gersema,
  Meyer~zum Alten~Borgloh, Schulze, Hartmann, Zenesini, and
  Ospelkaus}}]{Voges:2020}
\bibinfo{author}{\bibfnamefont{K.~K.} \bibnamefont{Voges}},
  \bibinfo{author}{\bibfnamefont{P.}~\bibnamefont{Gersema}},
  \bibinfo{author}{\bibfnamefont{M.}~\bibnamefont{Meyer~zum Alten~Borgloh}},
  \bibinfo{author}{\bibfnamefont{T.~A.} \bibnamefont{Schulze}},
  \bibinfo{author}{\bibfnamefont{T.}~\bibnamefont{Hartmann}},
  \bibinfo{author}{\bibfnamefont{A.}~\bibnamefont{Zenesini}}, \bibnamefont{and}
  \bibinfo{author}{\bibfnamefont{S.}~\bibnamefont{Ospelkaus}},
  \bibinfo{journal}{Phys. Rev. Lett.} \textbf{\bibinfo{volume}{125}},
  \bibinfo{pages}{083401} (\bibinfo{year}{2020}),
  \urlprefix\url{https://link.aps.org/doi/10.1103/PhysRevLett.125.083401}.

\bibitem[{\citenamefont{Shuman et~al.}(2010)\citenamefont{Shuman, Barry, and
  DeMille}}]{Shuman:2010}
\bibinfo{author}{\bibfnamefont{E.~S.} \bibnamefont{Shuman}},
  \bibinfo{author}{\bibfnamefont{J.~F.} \bibnamefont{Barry}}, \bibnamefont{and}
  \bibinfo{author}{\bibfnamefont{D.}~\bibnamefont{DeMille}},
  \bibinfo{journal}{Nature} \textbf{\bibinfo{volume}{467}},
  \bibinfo{pages}{820} (\bibinfo{year}{2010}).

\bibitem[{\citenamefont{Barry et~al.}(2014)\citenamefont{Barry, {McCarron},
  Norrgard, Steinecker, and {DeMille}}}]{Barry:2014}
\bibinfo{author}{\bibfnamefont{J.~F.} \bibnamefont{Barry}},
  \bibinfo{author}{\bibfnamefont{D.~J.} \bibnamefont{{McCarron}}},
  \bibinfo{author}{\bibfnamefont{E.~B.} \bibnamefont{Norrgard}},
  \bibinfo{author}{\bibfnamefont{M.~H.} \bibnamefont{Steinecker}},
  \bibnamefont{and}
  \bibinfo{author}{\bibfnamefont{D.}~\bibnamefont{{DeMille}}},
  \bibinfo{journal}{Nature} \textbf{\bibinfo{volume}{512}},
  \bibinfo{pages}{286} (\bibinfo{year}{2014}).

\bibitem[{\citenamefont{Truppe et~al.}(2017)\citenamefont{Truppe, Williams,
  Hambach, Caldwell, Fitch, Hinds, Sauer, and Tarbutt}}]{Truppe:2017}
\bibinfo{author}{\bibfnamefont{S.}~\bibnamefont{Truppe}},
  \bibinfo{author}{\bibfnamefont{H.~J.} \bibnamefont{Williams}},
  \bibinfo{author}{\bibfnamefont{M.}~\bibnamefont{Hambach}},
  \bibinfo{author}{\bibfnamefont{L.}~\bibnamefont{Caldwell}},
  \bibinfo{author}{\bibfnamefont{N.~J.} \bibnamefont{Fitch}},
  \bibinfo{author}{\bibfnamefont{E.~A.} \bibnamefont{Hinds}},
  \bibinfo{author}{\bibfnamefont{B.~E.} \bibnamefont{Sauer}}, \bibnamefont{and}
  \bibinfo{author}{\bibfnamefont{M.~R.} \bibnamefont{Tarbutt}},
  \bibinfo{journal}{Nature Physics} \textbf{\bibinfo{volume}{13}},
  \bibinfo{pages}{1173} (\bibinfo{year}{2017}).

\bibitem[{\citenamefont{Kozyryev et~al.}(2017)\citenamefont{Kozyryev, Baum,
  Matsuda, Augenbraun, Anderegg, Sedlack, and Doyle}}]{Kozyryev:2017}
\bibinfo{author}{\bibfnamefont{I.}~\bibnamefont{Kozyryev}},
  \bibinfo{author}{\bibfnamefont{L.}~\bibnamefont{Baum}},
  \bibinfo{author}{\bibfnamefont{K.}~\bibnamefont{Matsuda}},
  \bibinfo{author}{\bibfnamefont{B.~L.} \bibnamefont{Augenbraun}},
  \bibinfo{author}{\bibfnamefont{L.}~\bibnamefont{Anderegg}},
  \bibinfo{author}{\bibfnamefont{A.~P.} \bibnamefont{Sedlack}},
  \bibnamefont{and} \bibinfo{author}{\bibfnamefont{J.~M.} \bibnamefont{Doyle}},
  \bibinfo{journal}{Physical Review Letters} \textbf{\bibinfo{volume}{118}},
  \bibinfo{pages}{173201} (\bibinfo{year}{2017}).

\bibitem[{\citenamefont{Anderegg et~al.}(2018)\citenamefont{Anderegg,
  Augenbraun, Bao, Burchesky, Cheuk, Ketterle, and Doyle}}]{Anderegg:2018}
\bibinfo{author}{\bibfnamefont{L.}~\bibnamefont{Anderegg}},
  \bibinfo{author}{\bibfnamefont{B.~L.} \bibnamefont{Augenbraun}},
  \bibinfo{author}{\bibfnamefont{Y.}~\bibnamefont{Bao}},
  \bibinfo{author}{\bibfnamefont{S.}~\bibnamefont{Burchesky}},
  \bibinfo{author}{\bibfnamefont{L.~W.} \bibnamefont{Cheuk}},
  \bibinfo{author}{\bibfnamefont{W.}~\bibnamefont{Ketterle}}, \bibnamefont{and}
  \bibinfo{author}{\bibfnamefont{J.~M.} \bibnamefont{Doyle}},
  \bibinfo{journal}{Nature Physics} \textbf{\bibinfo{volume}{14}},
  \bibinfo{pages}{890} (\bibinfo{year}{2018}), ISSN \bibinfo{issn}{1745-2481},
  \urlprefix\url{https://doi.org/10.1038/s41567-018-0191-z}.

\bibitem[{\citenamefont{Collopy et~al.}(2018)\citenamefont{Collopy, Ding, Wu,
  Finneran, Anderegg, Augenbraun, Doyle, and Ye}}]{Collopy:2018}
\bibinfo{author}{\bibfnamefont{A.~L.} \bibnamefont{Collopy}},
  \bibinfo{author}{\bibfnamefont{S.}~\bibnamefont{Ding}},
  \bibinfo{author}{\bibfnamefont{Y.}~\bibnamefont{Wu}},
  \bibinfo{author}{\bibfnamefont{I.~A.} \bibnamefont{Finneran}},
  \bibinfo{author}{\bibfnamefont{L.}~\bibnamefont{Anderegg}},
  \bibinfo{author}{\bibfnamefont{B.~L.} \bibnamefont{Augenbraun}},
  \bibinfo{author}{\bibfnamefont{J.~M.} \bibnamefont{Doyle}}, \bibnamefont{and}
  \bibinfo{author}{\bibfnamefont{J.}~\bibnamefont{Ye}}, \bibinfo{journal}{Phys.
  Rev. Lett.} \textbf{\bibinfo{volume}{121}}, \bibinfo{pages}{213201}
  (\bibinfo{year}{2018}),
  \urlprefix\url{https://link.aps.org/doi/10.1103/PhysRevLett.121.213201}.

\bibitem[{\citenamefont{Ospelkaus
  et~al.}(2010{\natexlab{b}})\citenamefont{Ospelkaus, Ni, Qu\'em\'ener,
  Neyenhuis, Wang, de~Miranda, Bohn, Ye, and Jin}}]{Ospelkaus:2010b}
\bibinfo{author}{\bibfnamefont{S.}~\bibnamefont{Ospelkaus}},
  \bibinfo{author}{\bibfnamefont{K.-K.} \bibnamefont{Ni}},
  \bibinfo{author}{\bibfnamefont{G.}~\bibnamefont{Qu\'em\'ener}},
  \bibinfo{author}{\bibfnamefont{B.}~\bibnamefont{Neyenhuis}},
  \bibinfo{author}{\bibfnamefont{D.}~\bibnamefont{Wang}},
  \bibinfo{author}{\bibfnamefont{M.~H.~G.} \bibnamefont{de~Miranda}},
  \bibinfo{author}{\bibfnamefont{J.~L.} \bibnamefont{Bohn}},
  \bibinfo{author}{\bibfnamefont{J.}~\bibnamefont{Ye}}, \bibnamefont{and}
  \bibinfo{author}{\bibfnamefont{D.~S.} \bibnamefont{Jin}},
  \bibinfo{journal}{Phys. Rev. Lett.} \textbf{\bibinfo{volume}{104}},
  \bibinfo{pages}{030402} (\bibinfo{year}{2010}{\natexlab{b}}),
  \urlprefix\url{https://link.aps.org/doi/10.1103/PhysRevLett.104.030402}.

\bibitem[{\citenamefont{Yan et~al.}(2013)\citenamefont{Yan, Moses, Gadway,
  Covey, Hazzard, Rey, Jin, and Ye}}]{Yan:2013}
\bibinfo{author}{\bibfnamefont{B.}~\bibnamefont{Yan}},
  \bibinfo{author}{\bibfnamefont{S.~A.} \bibnamefont{Moses}},
  \bibinfo{author}{\bibfnamefont{B.}~\bibnamefont{Gadway}},
  \bibinfo{author}{\bibfnamefont{J.~P.} \bibnamefont{Covey}},
  \bibinfo{author}{\bibfnamefont{K.~R.} \bibnamefont{Hazzard}},
  \bibinfo{author}{\bibfnamefont{A.~M.} \bibnamefont{Rey}},
  \bibinfo{author}{\bibfnamefont{D.~S.} \bibnamefont{Jin}}, \bibnamefont{and}
  \bibinfo{author}{\bibfnamefont{J.}~\bibnamefont{Ye}},
  \bibinfo{journal}{Nature} \textbf{\bibinfo{volume}{501}},
  \bibinfo{pages}{521} (\bibinfo{year}{2013}), ISSN \bibinfo{issn}{00280836}.

\bibitem[{\citenamefont{Gregory et~al.}(2016)\citenamefont{Gregory, Aldegunde,
  Hutson, and Cornish}}]{Gregory:2016}
\bibinfo{author}{\bibfnamefont{P.~D.} \bibnamefont{Gregory}},
  \bibinfo{author}{\bibfnamefont{J.}~\bibnamefont{Aldegunde}},
  \bibinfo{author}{\bibfnamefont{J.~M.} \bibnamefont{Hutson}},
  \bibnamefont{and} \bibinfo{author}{\bibfnamefont{S.~L.}
  \bibnamefont{Cornish}}, \bibinfo{journal}{Phys. Rev. A}
  \textbf{\bibinfo{volume}{94}}, \bibinfo{pages}{041403}
  (\bibinfo{year}{2016}),
  \urlprefix\url{https://link.aps.org/doi/10.1103/PhysRevA.94.041403}.

\bibitem[{\citenamefont{Will et~al.}(2016)\citenamefont{Will, Park, Yan, Loh,
  and Zwierlein}}]{Will:2016}
\bibinfo{author}{\bibfnamefont{S.~A.} \bibnamefont{Will}},
  \bibinfo{author}{\bibfnamefont{J.~W.} \bibnamefont{Park}},
  \bibinfo{author}{\bibfnamefont{Z.~Z.} \bibnamefont{Yan}},
  \bibinfo{author}{\bibfnamefont{H.}~\bibnamefont{Loh}}, \bibnamefont{and}
  \bibinfo{author}{\bibfnamefont{M.~W.} \bibnamefont{Zwierlein}},
  \bibinfo{journal}{Phys. Rev. Lett.} \textbf{\bibinfo{volume}{116}},
  \bibinfo{pages}{225306} (\bibinfo{year}{2016}),
  \urlprefix\url{https://link.aps.org/doi/10.1103/PhysRevLett.116.225306}.

\bibitem[{\citenamefont{Guo et~al.}(2018)\citenamefont{Guo, Ye, He,
  Qu\'em\'ener, and Wang}}]{Guo:2018}
\bibinfo{author}{\bibfnamefont{M.}~\bibnamefont{Guo}},
  \bibinfo{author}{\bibfnamefont{X.}~\bibnamefont{Ye}},
  \bibinfo{author}{\bibfnamefont{J.}~\bibnamefont{He}},
  \bibinfo{author}{\bibfnamefont{G.}~\bibnamefont{Qu\'em\'ener}},
  \bibnamefont{and} \bibinfo{author}{\bibfnamefont{D.}~\bibnamefont{Wang}},
  \bibinfo{journal}{Phys. Rev. A} \textbf{\bibinfo{volume}{97}},
  \bibinfo{pages}{020501} (\bibinfo{year}{2018}),
  \urlprefix\url{https://link.aps.org/doi/10.1103/PhysRevA.97.020501}.

\bibitem[{\citenamefont{Blackmore et~al.}(2020)\citenamefont{Blackmore,
  Gregory, Bromley, and Cornish}}]{Blackmore:2020}
\bibinfo{author}{\bibfnamefont{J.~A.} \bibnamefont{Blackmore}},
  \bibinfo{author}{\bibfnamefont{P.~D.} \bibnamefont{Gregory}},
  \bibinfo{author}{\bibfnamefont{S.~L.} \bibnamefont{Bromley}},
  \bibnamefont{and} \bibinfo{author}{\bibfnamefont{S.~L.}
  \bibnamefont{Cornish}}, \bibinfo{journal}{Phys. Chem. Chem. Phys.}
  pp.~\bibinfo{pages}{--} (\bibinfo{year}{2020}),
  \urlprefix\url{http://dx.doi.org/10.1039/D0CP04651E}.

\bibitem[{\citenamefont{Gorshkov
  et~al.}(2011{\natexlab{a}})\citenamefont{Gorshkov, Manmana, Chen, Ye, Demler,
  Lukin, and Rey}}]{Gorshkov:2011}
\bibinfo{author}{\bibfnamefont{A.~V.} \bibnamefont{Gorshkov}},
  \bibinfo{author}{\bibfnamefont{S.~R.} \bibnamefont{Manmana}},
  \bibinfo{author}{\bibfnamefont{G.}~\bibnamefont{Chen}},
  \bibinfo{author}{\bibfnamefont{J.}~\bibnamefont{Ye}},
  \bibinfo{author}{\bibfnamefont{E.}~\bibnamefont{Demler}},
  \bibinfo{author}{\bibfnamefont{M.~D.} \bibnamefont{Lukin}}, \bibnamefont{and}
  \bibinfo{author}{\bibfnamefont{A.~M.} \bibnamefont{Rey}},
  \bibinfo{journal}{Physical Review Letters} \textbf{\bibinfo{volume}{107}},
  \bibinfo{pages}{115301} (\bibinfo{year}{2011}{\natexlab{a}}),
  \urlprefix\url{https://doi.org/10.1103%2Fphysrevlett.107.115301}.

\bibitem[{\citenamefont{Gorshkov
  et~al.}(2011{\natexlab{b}})\citenamefont{Gorshkov, Manmana, Chen, Demler,
  Lukin, and Rey}}]{Gorshkov:2011b}
\bibinfo{author}{\bibfnamefont{A.~V.} \bibnamefont{Gorshkov}},
  \bibinfo{author}{\bibfnamefont{S.~R.} \bibnamefont{Manmana}},
  \bibinfo{author}{\bibfnamefont{G.}~\bibnamefont{Chen}},
  \bibinfo{author}{\bibfnamefont{E.}~\bibnamefont{Demler}},
  \bibinfo{author}{\bibfnamefont{M.~D.} \bibnamefont{Lukin}}, \bibnamefont{and}
  \bibinfo{author}{\bibfnamefont{A.~M.} \bibnamefont{Rey}},
  \bibinfo{journal}{Physical Review A} \textbf{\bibinfo{volume}{84}},
  \bibinfo{pages}{033619} (\bibinfo{year}{2011}{\natexlab{b}}),
  \urlprefix\url{https://doi.org/10.1103%2Fphysreva.84.033619}.

\bibitem[{\citenamefont{Hazzard et~al.}(2013)\citenamefont{Hazzard, Manmana,
  Foss-Feig, and Rey}}]{Hazzard:2013}
\bibinfo{author}{\bibfnamefont{K.~R.~A.} \bibnamefont{Hazzard}},
  \bibinfo{author}{\bibfnamefont{S.~R.} \bibnamefont{Manmana}},
  \bibinfo{author}{\bibfnamefont{M.}~\bibnamefont{Foss-Feig}},
  \bibnamefont{and} \bibinfo{author}{\bibfnamefont{A.~M.} \bibnamefont{Rey}},
  \bibinfo{journal}{Physical Review Letters} \textbf{\bibinfo{volume}{110}},
  \bibinfo{pages}{075301} (\bibinfo{year}{2013}),
  \urlprefix\url{https://doi.org/10.1103%2Fphysrevlett.110.075301}.

\bibitem[{\citenamefont{Moses et~al.}(2015)\citenamefont{Moses, Covey,
  Miecnikowski, Yan, Gadway, Ye, and Jin}}]{Moses:2015}
\bibinfo{author}{\bibfnamefont{S.~A.} \bibnamefont{Moses}},
  \bibinfo{author}{\bibfnamefont{J.~P.} \bibnamefont{Covey}},
  \bibinfo{author}{\bibfnamefont{M.~T.} \bibnamefont{Miecnikowski}},
  \bibinfo{author}{\bibfnamefont{B.}~\bibnamefont{Yan}},
  \bibinfo{author}{\bibfnamefont{B.}~\bibnamefont{Gadway}},
  \bibinfo{author}{\bibfnamefont{J.}~\bibnamefont{Ye}}, \bibnamefont{and}
  \bibinfo{author}{\bibfnamefont{D.~S.} \bibnamefont{Jin}},
  \bibinfo{journal}{Science} \textbf{\bibinfo{volume}{350}},
  \bibinfo{pages}{659} (\bibinfo{year}{2015}), ISSN \bibinfo{issn}{0036-8075},
  \urlprefix\url{https://science.sciencemag.org/content/350/6261/659}.

\bibitem[{\citenamefont{Reichs\"ollner
  et~al.}(2017)\citenamefont{Reichs\"ollner, Schindewolf, Takekoshi, Grimm, and
  N\"agerl}}]{Reichsollner:2017}
\bibinfo{author}{\bibfnamefont{L.}~\bibnamefont{Reichs\"ollner}},
  \bibinfo{author}{\bibfnamefont{A.}~\bibnamefont{Schindewolf}},
  \bibinfo{author}{\bibfnamefont{T.}~\bibnamefont{Takekoshi}},
  \bibinfo{author}{\bibfnamefont{R.}~\bibnamefont{Grimm}}, \bibnamefont{and}
  \bibinfo{author}{\bibfnamefont{H.-C.} \bibnamefont{N\"agerl}},
  \bibinfo{journal}{Phys. Rev. Lett.} \textbf{\bibinfo{volume}{118}},
  \bibinfo{pages}{073201} (\bibinfo{year}{2017}),
  \urlprefix\url{https://link.aps.org/doi/10.1103/PhysRevLett.118.073201}.

\bibitem[{\citenamefont{Liu et~al.}(2019)\citenamefont{Liu, Hood, Yu, Zhang,
  Wang, Lin, Rosenband, and Ni}}]{Liu:2019}
\bibinfo{author}{\bibfnamefont{L.~R.} \bibnamefont{Liu}},
  \bibinfo{author}{\bibfnamefont{J.~D.} \bibnamefont{Hood}},
  \bibinfo{author}{\bibfnamefont{Y.}~\bibnamefont{Yu}},
  \bibinfo{author}{\bibfnamefont{J.~T.} \bibnamefont{Zhang}},
  \bibinfo{author}{\bibfnamefont{K.}~\bibnamefont{Wang}},
  \bibinfo{author}{\bibfnamefont{Y.-W.} \bibnamefont{Lin}},
  \bibinfo{author}{\bibfnamefont{T.}~\bibnamefont{Rosenband}},
  \bibnamefont{and} \bibinfo{author}{\bibfnamefont{K.-K.} \bibnamefont{Ni}},
  \bibinfo{journal}{Phys. Rev. X} \textbf{\bibinfo{volume}{9}},
  \bibinfo{pages}{021039} (\bibinfo{year}{2019}),
  \urlprefix\url{https://link.aps.org/doi/10.1103/PhysRevX.9.021039}.

\bibitem[{\citenamefont{Anderegg et~al.}(2019)\citenamefont{Anderegg, Cheuk,
  Bao, Burchesky, Ketterle, Ni, and Doyle}}]{Anderegg:2019}
\bibinfo{author}{\bibfnamefont{L.}~\bibnamefont{Anderegg}},
  \bibinfo{author}{\bibfnamefont{L.~W.} \bibnamefont{Cheuk}},
  \bibinfo{author}{\bibfnamefont{Y.}~\bibnamefont{Bao}},
  \bibinfo{author}{\bibfnamefont{S.}~\bibnamefont{Burchesky}},
  \bibinfo{author}{\bibfnamefont{W.}~\bibnamefont{Ketterle}},
  \bibinfo{author}{\bibfnamefont{K.-K.} \bibnamefont{Ni}}, \bibnamefont{and}
  \bibinfo{author}{\bibfnamefont{J.~M.} \bibnamefont{Doyle}},
  \bibinfo{journal}{Science} \textbf{\bibinfo{volume}{365}},
  \bibinfo{pages}{1156} (\bibinfo{year}{2019}), ISSN \bibinfo{issn}{0036-8075},
  \urlprefix\url{https://science.sciencemag.org/content/365/6458/1156}.

\bibitem[{\citenamefont{Kotochigova and Tiesinga}(2006)}]{Kotochigova:2006}
\bibinfo{author}{\bibfnamefont{S.}~\bibnamefont{Kotochigova}} \bibnamefont{and}
  \bibinfo{author}{\bibfnamefont{E.}~\bibnamefont{Tiesinga}},
  \bibinfo{journal}{Phys. Rev. A} \textbf{\bibinfo{volume}{73}},
  \bibinfo{pages}{041405} (\bibinfo{year}{2006}),
  \urlprefix\url{https://link.aps.org/doi/10.1103/PhysRevA.73.041405}.

\bibitem[{\citenamefont{Vexiau et~al.}(2017)\citenamefont{Vexiau, Borsalino,
  Lepers, Orbán, Aymar, Dulieu, and Bouloufa-Maafa}}]{Vexiau:2017}
\bibinfo{author}{\bibfnamefont{R.}~\bibnamefont{Vexiau}},
  \bibinfo{author}{\bibfnamefont{D.}~\bibnamefont{Borsalino}},
  \bibinfo{author}{\bibfnamefont{M.}~\bibnamefont{Lepers}},
  \bibinfo{author}{\bibfnamefont{A.}~\bibnamefont{Orbán}},
  \bibinfo{author}{\bibfnamefont{M.}~\bibnamefont{Aymar}},
  \bibinfo{author}{\bibfnamefont{O.}~\bibnamefont{Dulieu}}, \bibnamefont{and}
  \bibinfo{author}{\bibfnamefont{N.}~\bibnamefont{Bouloufa-Maafa}},
  \bibinfo{journal}{International Reviews in Physical Chemistry}
  \textbf{\bibinfo{volume}{36}}, \bibinfo{pages}{709} (\bibinfo{year}{2017}),
  \eprint{https://doi.org/10.1080/0144235X.2017.1351821},
  \urlprefix\url{https://doi.org/10.1080/0144235X.2017.1351821}.

\bibitem[{\citenamefont{Li et~al.}(2017)\citenamefont{Li, Petrov, Makrides,
  Tiesinga, and Kotochigova}}]{Li:2017}
\bibinfo{author}{\bibfnamefont{M.}~\bibnamefont{Li}},
  \bibinfo{author}{\bibfnamefont{A.}~\bibnamefont{Petrov}},
  \bibinfo{author}{\bibfnamefont{C.}~\bibnamefont{Makrides}},
  \bibinfo{author}{\bibfnamefont{E.}~\bibnamefont{Tiesinga}}, \bibnamefont{and}
  \bibinfo{author}{\bibfnamefont{S.}~\bibnamefont{Kotochigova}},
  \bibinfo{journal}{Phys. Rev. A} \textbf{\bibinfo{volume}{95}},
  \bibinfo{pages}{063422} (\bibinfo{year}{2017}),
  \urlprefix\url{https://link.aps.org/doi/10.1103/PhysRevA.95.063422}.

\bibitem[{\citenamefont{See\ss{}elberg
  et~al.}(2018)\citenamefont{See\ss{}elberg, Luo, Li, Bause, Kotochigova,
  Bloch, and Gohle}}]{Seesselberg:2018}
\bibinfo{author}{\bibfnamefont{F.}~\bibnamefont{See\ss{}elberg}},
  \bibinfo{author}{\bibfnamefont{X.-Y.} \bibnamefont{Luo}},
  \bibinfo{author}{\bibfnamefont{M.}~\bibnamefont{Li}},
  \bibinfo{author}{\bibfnamefont{R.}~\bibnamefont{Bause}},
  \bibinfo{author}{\bibfnamefont{S.}~\bibnamefont{Kotochigova}},
  \bibinfo{author}{\bibfnamefont{I.}~\bibnamefont{Bloch}}, \bibnamefont{and}
  \bibinfo{author}{\bibfnamefont{C.}~\bibnamefont{Gohle}},
  \bibinfo{journal}{Phys. Rev. Lett.} \textbf{\bibinfo{volume}{121}},
  \bibinfo{pages}{253401} (\bibinfo{year}{2018}),
  \urlprefix\url{https://link.aps.org/doi/10.1103/PhysRevLett.121.253401}.

\bibitem[{\citenamefont{Neyenhuis et~al.}(2012)\citenamefont{Neyenhuis, Yan,
  Moses, Covey, Chotia, Petrov, Kotochigova, Ye, and Jin}}]{Neyenhuis:2012}
\bibinfo{author}{\bibfnamefont{B.}~\bibnamefont{Neyenhuis}},
  \bibinfo{author}{\bibfnamefont{B.}~\bibnamefont{Yan}},
  \bibinfo{author}{\bibfnamefont{S.~A.} \bibnamefont{Moses}},
  \bibinfo{author}{\bibfnamefont{J.~P.} \bibnamefont{Covey}},
  \bibinfo{author}{\bibfnamefont{A.}~\bibnamefont{Chotia}},
  \bibinfo{author}{\bibfnamefont{A.}~\bibnamefont{Petrov}},
  \bibinfo{author}{\bibfnamefont{S.}~\bibnamefont{Kotochigova}},
  \bibinfo{author}{\bibfnamefont{J.}~\bibnamefont{Ye}}, \bibnamefont{and}
  \bibinfo{author}{\bibfnamefont{D.~S.} \bibnamefont{Jin}},
  \bibinfo{journal}{Physical Review Letters} \textbf{\bibinfo{volume}{109}},
  \bibinfo{pages}{230403} (\bibinfo{year}{2012}),
  \urlprefix\url{https://link.aps.org/doi/10.1103/PhysRevLett.109.230403}.

\bibitem[{\citenamefont{Gregory et~al.}(2017)\citenamefont{Gregory, Blackmore,
  Aldegunde, Hutson, and Cornish}}]{Gregory:2017}
\bibinfo{author}{\bibfnamefont{P.~D.} \bibnamefont{Gregory}},
  \bibinfo{author}{\bibfnamefont{J.~A.} \bibnamefont{Blackmore}},
  \bibinfo{author}{\bibfnamefont{J.}~\bibnamefont{Aldegunde}},
  \bibinfo{author}{\bibfnamefont{J.~M.} \bibnamefont{Hutson}},
  \bibnamefont{and} \bibinfo{author}{\bibfnamefont{S.~L.}
  \bibnamefont{Cornish}}, \bibinfo{journal}{Physical Review A}
  \textbf{\bibinfo{volume}{96}}, \bibinfo{pages}{021402(R)}
  (\bibinfo{year}{2017}).

\bibitem[{\citenamefont{Blackmore et~al.}(2018)\citenamefont{Blackmore,
  Caldwell, Gregory, Bridge, Sawant, Aldegunde, Mur-Petit, Jaksch, Hutson,
  Sauer et~al.}}]{Blackmore:2018}
\bibinfo{author}{\bibfnamefont{J.~A.} \bibnamefont{Blackmore}},
  \bibinfo{author}{\bibfnamefont{L.}~\bibnamefont{Caldwell}},
  \bibinfo{author}{\bibfnamefont{P.~D.} \bibnamefont{Gregory}},
  \bibinfo{author}{\bibfnamefont{E.~M.} \bibnamefont{Bridge}},
  \bibinfo{author}{\bibfnamefont{R.}~\bibnamefont{Sawant}},
  \bibinfo{author}{\bibfnamefont{J.}~\bibnamefont{Aldegunde}},
  \bibinfo{author}{\bibfnamefont{J.}~\bibnamefont{Mur-Petit}},
  \bibinfo{author}{\bibfnamefont{D.}~\bibnamefont{Jaksch}},
  \bibinfo{author}{\bibfnamefont{J.~M.} \bibnamefont{Hutson}},
  \bibinfo{author}{\bibfnamefont{B.~E.} \bibnamefont{Sauer}},
  \bibnamefont{et~al.}, \bibinfo{journal}{Quantum Science and Technology}
  \textbf{\bibinfo{volume}{4}}, \bibinfo{pages}{014010} (\bibinfo{year}{2018}),
  \urlprefix\url{https://doi.org/10.1088%2F2058-9565%2Faaee35}.

\bibitem[{\citenamefont{Kotochigova and DeMille}(2010)}]{Kotochigova:2010}
\bibinfo{author}{\bibfnamefont{S.}~\bibnamefont{Kotochigova}} \bibnamefont{and}
  \bibinfo{author}{\bibfnamefont{D.}~\bibnamefont{DeMille}},
  \bibinfo{journal}{Physical Review A} \textbf{\bibinfo{volume}{82}},
  \bibinfo{pages}{063421} (\bibinfo{year}{2010}),
  \urlprefix\url{https://link.aps.org/doi/10.1103/PhysRevA.82.063421}.

\bibitem[{\citenamefont{Katori et~al.}(2003)\citenamefont{Katori, Takamoto,
  Pal'chikov, and Ovsiannikov}}]{Katori:2003}
\bibinfo{author}{\bibfnamefont{H.}~\bibnamefont{Katori}},
  \bibinfo{author}{\bibfnamefont{M.}~\bibnamefont{Takamoto}},
  \bibinfo{author}{\bibfnamefont{V.~G.} \bibnamefont{Pal'chikov}},
  \bibnamefont{and} \bibinfo{author}{\bibfnamefont{V.~D.}
  \bibnamefont{Ovsiannikov}}, \bibinfo{journal}{Phys. Rev. Lett.}
  \textbf{\bibinfo{volume}{91}}, \bibinfo{pages}{173005}
  (\bibinfo{year}{2003}),
  \urlprefix\url{https://link.aps.org/doi/10.1103/PhysRevLett.91.173005}.

\bibitem[{\citenamefont{Ye et~al.}(2008)\citenamefont{Ye, Kimble, and
  Katori}}]{Ye:2008}
\bibinfo{author}{\bibfnamefont{J.}~\bibnamefont{Ye}},
  \bibinfo{author}{\bibfnamefont{H.~J.} \bibnamefont{Kimble}},
  \bibnamefont{and} \bibinfo{author}{\bibfnamefont{H.}~\bibnamefont{Katori}},
  \bibinfo{journal}{Science} \textbf{\bibinfo{volume}{320}},
  \bibinfo{pages}{1734} (\bibinfo{year}{2008}), ISSN \bibinfo{issn}{0036-8075},
  \urlprefix\url{https://science.sciencemag.org/content/320/5884/1734}.

\bibitem[{\citenamefont{Kondov et~al.}(2019)\citenamefont{Kondov, Lee, Leung,
  Liedl, Majewska, Moszynski, and Zelevinsky}}]{Kondov:2019}
\bibinfo{author}{\bibfnamefont{S.~S.} \bibnamefont{Kondov}},
  \bibinfo{author}{\bibfnamefont{C.~H.} \bibnamefont{Lee}},
  \bibinfo{author}{\bibfnamefont{K.~H.} \bibnamefont{Leung}},
  \bibinfo{author}{\bibfnamefont{C.}~\bibnamefont{Liedl}},
  \bibinfo{author}{\bibfnamefont{I.}~\bibnamefont{Majewska}},
  \bibinfo{author}{\bibfnamefont{R.}~\bibnamefont{Moszynski}},
  \bibnamefont{and}
  \bibinfo{author}{\bibfnamefont{T.}~\bibnamefont{Zelevinsky}},
  \bibinfo{journal}{Nature Physics} \textbf{\bibinfo{volume}{15}},
  \bibinfo{pages}{1118} (\bibinfo{year}{2019}).

\bibitem[{\citenamefont{Bause et~al.}(2020)\citenamefont{Bause, Li,
  Schindewolf, Chen, Duda, Kotochigova, Bloch, and Luo}}]{Bause:2019}
\bibinfo{author}{\bibfnamefont{R.}~\bibnamefont{Bause}},
  \bibinfo{author}{\bibfnamefont{M.}~\bibnamefont{Li}},
  \bibinfo{author}{\bibfnamefont{A.}~\bibnamefont{Schindewolf}},
  \bibinfo{author}{\bibfnamefont{X.-Y.} \bibnamefont{Chen}},
  \bibinfo{author}{\bibfnamefont{M.}~\bibnamefont{Duda}},
  \bibinfo{author}{\bibfnamefont{S.}~\bibnamefont{Kotochigova}},
  \bibinfo{author}{\bibfnamefont{I.}~\bibnamefont{Bloch}}, \bibnamefont{and}
  \bibinfo{author}{\bibfnamefont{X.-Y.} \bibnamefont{Luo}},
  \bibinfo{journal}{Phys. Rev. Lett.} \textbf{\bibinfo{volume}{125}},
  \bibinfo{pages}{023201} (\bibinfo{year}{2020}),
  \urlprefix\url{https://link.aps.org/doi/10.1103/PhysRevLett.125.023201}.

\bibitem[{\citenamefont{Sundar et~al.}(2018)\citenamefont{Sundar, Gadway, and
  Hazzard}}]{Sundar:2018}
\bibinfo{author}{\bibfnamefont{B.}~\bibnamefont{Sundar}},
  \bibinfo{author}{\bibfnamefont{B.}~\bibnamefont{Gadway}}, \bibnamefont{and}
  \bibinfo{author}{\bibfnamefont{K.~R.} \bibnamefont{Hazzard}},
  \bibinfo{journal}{Scientific Reports} \textbf{\bibinfo{volume}{8}},
  \bibinfo{pages}{1} (\bibinfo{year}{2018}), ISSN \bibinfo{issn}{20452322}.

\bibitem[{\citenamefont{Petrov et~al.}(2013)\citenamefont{Petrov, Makrides, and
  Kotochigova}}]{APetrov:2013}
\bibinfo{author}{\bibfnamefont{A.}~\bibnamefont{Petrov}},
  \bibinfo{author}{\bibfnamefont{C.}~\bibnamefont{Makrides}}, \bibnamefont{and}
  \bibinfo{author}{\bibfnamefont{S.}~\bibnamefont{Kotochigova}},
  \bibinfo{journal}{Mol. Phys.} \textbf{\bibinfo{volume}{111}},
  \bibinfo{pages}{1731} (\bibinfo{year}{2013}).

\bibitem[{\citenamefont{Aldegunde et~al.}(2008)\citenamefont{Aldegunde,
  Rivington, \ifmmode~\dot{Z}\else \.{Z}\fi{}uchowski, and
  Hutson}}]{Aldegunde:2008}
\bibinfo{author}{\bibfnamefont{J.}~\bibnamefont{Aldegunde}},
  \bibinfo{author}{\bibfnamefont{B.~A.} \bibnamefont{Rivington}},
  \bibinfo{author}{\bibfnamefont{P.~S.} \bibnamefont{\ifmmode~\dot{Z}\else
  \.{Z}\fi{}uchowski}}, \bibnamefont{and} \bibinfo{author}{\bibfnamefont{J.~M.}
  \bibnamefont{Hutson}}, \bibinfo{journal}{Phys. Rev. A}
  \textbf{\bibinfo{volume}{78}}, \bibinfo{pages}{033434}
  (\bibinfo{year}{2008}),
  \urlprefix\url{https://link.aps.org/doi/10.1103/PhysRevA.78.033434}.

\bibitem[{\citenamefont{Docenko et~al.}(2010)\citenamefont{Docenko, Tamanis,
  Ferber, Bergeman, Kotochigova, Stolyarov, de~Faria~Nogueira, and
  Fellows}}]{Docenko:2010}
\bibinfo{author}{\bibfnamefont{O.}~\bibnamefont{Docenko}},
  \bibinfo{author}{\bibfnamefont{M.}~\bibnamefont{Tamanis}},
  \bibinfo{author}{\bibfnamefont{R.}~\bibnamefont{Ferber}},
  \bibinfo{author}{\bibfnamefont{T.}~\bibnamefont{Bergeman}},
  \bibinfo{author}{\bibfnamefont{S.}~\bibnamefont{Kotochigova}},
  \bibinfo{author}{\bibfnamefont{A.~V.} \bibnamefont{Stolyarov}},
  \bibinfo{author}{\bibfnamefont{A.}~\bibnamefont{de~Faria~Nogueira}},
  \bibnamefont{and} \bibinfo{author}{\bibfnamefont{C.~E.}
  \bibnamefont{Fellows}}, \bibinfo{journal}{Phys. Rev. A}
  \textbf{\bibinfo{volume}{81}}, \bibinfo{pages}{042511}
  (\bibinfo{year}{2010}),
  \urlprefix\url{https://link.aps.org/doi/10.1103/PhysRevA.81.042511}.

\bibitem[{\citenamefont{Docenko et~al.}(2011)\citenamefont{Docenko, Tamanis,
  Ferber, Kn\"ockel, and Tiemann}}]{Docenko:2011}
\bibinfo{author}{\bibfnamefont{O.}~\bibnamefont{Docenko}},
  \bibinfo{author}{\bibfnamefont{M.}~\bibnamefont{Tamanis}},
  \bibinfo{author}{\bibfnamefont{R.}~\bibnamefont{Ferber}},
  \bibinfo{author}{\bibfnamefont{H.}~\bibnamefont{Kn\"ockel}},
  \bibnamefont{and} \bibinfo{author}{\bibfnamefont{E.}~\bibnamefont{Tiemann}},
  \bibinfo{journal}{Phys. Rev. A} \textbf{\bibinfo{volume}{83}},
  \bibinfo{pages}{052519} (\bibinfo{year}{2011}),
  \urlprefix\url{https://link.aps.org/doi/10.1}.

\bibitem[{\citenamefont{Raki\'c et~al.}(2016)\citenamefont{Raki\'c, Beuc,
  Bouloufa-Maafa, Dulieu, Vexiau, Pichler, and Skenderovi\'c}}]{Mario:2016}
\bibinfo{author}{\bibfnamefont{M.}~\bibnamefont{Raki\'c}},
  \bibinfo{author}{\bibfnamefont{R.}~\bibnamefont{Beuc}},
  \bibinfo{author}{\bibfnamefont{N.}~\bibnamefont{Bouloufa-Maafa}},
  \bibinfo{author}{\bibfnamefont{O.}~\bibnamefont{Dulieu}},
  \bibinfo{author}{\bibfnamefont{R.}~\bibnamefont{Vexiau}},
  \bibinfo{author}{\bibfnamefont{G.}~\bibnamefont{Pichler}}, \bibnamefont{and}
  \bibinfo{author}{\bibfnamefont{H.}~\bibnamefont{Skenderovi\'c}},
  \bibinfo{journal}{J. Chem. Phys.} \textbf{\bibinfo{volume}{144}},
  \bibinfo{pages}{204310} (\bibinfo{year}{2016}),
  \urlprefix\url{https://doi.org/10.1063/1.4952758}.

\bibitem[{\citenamefont{Chotia et~al.}(2012)\citenamefont{Chotia, Neyenhuis,
  Moses, Yan, Covey, M., Rey, Jin, and Ye}}]{Chotia:2012}
\bibinfo{author}{\bibfnamefont{A.}~\bibnamefont{Chotia}},
  \bibinfo{author}{\bibfnamefont{B.}~\bibnamefont{Neyenhuis}},
  \bibinfo{author}{\bibfnamefont{S.~A.} \bibnamefont{Moses}},
  \bibinfo{author}{\bibfnamefont{B.}~\bibnamefont{Yan}},
  \bibinfo{author}{\bibfnamefont{J.~P.} \bibnamefont{Covey}},
  \bibinfo{author}{\bibfnamefont{F.-F.} \bibnamefont{M.}},
  \bibinfo{author}{\bibfnamefont{A.~M.} \bibnamefont{Rey}},
  \bibinfo{author}{\bibfnamefont{D.~S.} \bibnamefont{Jin}}, \bibnamefont{and}
  \bibinfo{author}{\bibfnamefont{J.}~\bibnamefont{Ye}},
  \bibinfo{journal}{Physical Review Letters} \textbf{\bibinfo{volume}{108}},
  \bibinfo{pages}{080405} (\bibinfo{year}{2012}).

\bibitem[{\citenamefont{Schuster et~al.}(2019)\citenamefont{Schuster, Flicker,
  Li, Kotochigova, Moore, Ye, and Yao}}]{Schuster:2019}
\bibinfo{author}{\bibfnamefont{T.}~\bibnamefont{Schuster}},
  \bibinfo{author}{\bibfnamefont{F.}~\bibnamefont{Flicker}},
  \bibinfo{author}{\bibfnamefont{M.}~\bibnamefont{Li}},
  \bibinfo{author}{\bibfnamefont{S.}~\bibnamefont{Kotochigova}},
  \bibinfo{author}{\bibfnamefont{J.~E.} \bibnamefont{Moore}},
  \bibinfo{author}{\bibfnamefont{J.}~\bibnamefont{Ye}}, \bibnamefont{and}
  \bibinfo{author}{\bibfnamefont{N.~Y.} \bibnamefont{Yao}},
  \bibinfo{journal}{arXiv:1901.08597}  (\bibinfo{year}{2019}),
  \urlprefix\url{https://arxiv.org/abs/1901.08597}.

\end{thebibliography}
\end{document}